\documentclass[a4paper,fleqn]{cas-dc}
\usepackage[numbers]{natbib}
\usepackage{eurosym,makecell,comment}
\usepackage{multirow}
\usepackage{threeparttable}
\usepackage{framed}
\usepackage{hyperref}
\usepackage{url}
\usepackage{verbatim}
\usepackage{graphicx}
\usepackage{subcaption}
\usepackage{float}
\usepackage{xcolor}
\usepackage{microtype}
\usepackage{colortbl}
\usepackage{amsmath,amssymb,amsfonts}
\usepackage{tikz}
\usepackage{pgfplots}
\usepackage{listings}

\usepackage{soul}
\sethlcolor{yellow!50}

\usepackage{pifont}
\usepackage{algorithm}
\usepackage{algpseudocode}
\usepackage{balance}

\pgfplotsset{compat=1.18}

\newcommand{\ie}{\textit{i.e., }}  

\def \1{\textit{(i)}}
\def \2{\textit{(ii)}}
\def \3{\textit{(iii)}}
\def \4{\textit{(iv)}}
\def \5{\textit{(v)}}

\usepackage{etoolbox}
\usepackage{soul} 

\newtoggle{finalPaper}

\setstcolor{blue}

\toggletrue{finalPaper} 

\iftoggle{finalPaper} {

	\newcommand{\rmvtxt}[1]{}}
{

	\newcommand{\rmvtxt}[1]{\st{#1}}}

\newtoggle{removeItalic}
\togglefalse{removeItalic} 
\iftoggle{removeItalic} {
    \renewcommand{\textit}[1]{#1}
}

\newcommand{\solution}{\textit{GreenDFL}}

\begin{document}
\let\WriteBookmarks\relax
\def\floatpagepagefraction{1}
\def\textpagefraction{.001}
\shorttitle{Assessing the Sustainability of Decentralized Federated Learning Systems}
\shortauthors{Feng et~al.}

\title[mode = title]{\textit{GreenDFL}: a Framework for Assessing the Sustainability of Decentralized Federated Learning Systems}

\author[1]{Chao Feng}[orcid=0000-0002-0672-1090]
\author[1,2]{Alberto {Huertas Celdrán}}[orcid=0000-0001-7125-1710]
\author[1]{Xi Cheng} []
\author[2]{Gérôme Bovet} [orcid=0000-0002-4534-3483]
\author[3]{Burkhard Stiller}[orcid=0000-0002-7461-7463]

\address[1]{Communication Systems Group CSG, Department of Informatics IfI, University of Zurich UZH, CH---8050 Zürich, Switzerland}

\address[2]{Department of Information and Communications Engineering, University of Murcia, 30100--Murcia, Spain}

\address[3]{Cyber-Defence Campus within armasuisse Science \& Technology, CH---3602 Thun, Switzerland}

\cortext[cor1]{Corresponding author.
Email address: cfeng@ifi.uzh.ch (C. Feng)}

\maketitle

\begin{abstract}
\paragraph{\textbf{Context:}} Decentralized Federated Learning (DFL) is an emerging paradigm that enables collaborative model training without centralized data and model aggregation, enhancing privacy and resilience. However, its sustainability remains underexplored, as energy consumption and carbon emissions vary across different system configurations. Understanding the environmental impact of DFL is crucial for optimizing its design and deployment.\\
\textbf{Objective:} This work aims to develop a comprehensive and operational framework for assessing the sustainability of DFL systems. To address it, this work provides a systematic method for quantifying energy consumption and carbon emissions, offering insights into improving the sustainability of DFL.\\
\textbf{Methods:} This work proposes \solution{}, a fully implementable framework that has been integrated into a real-world DFL platform. \solution{} systematically analyzes the impact of various factors, including hardware accelerators, model architecture, communication medium, data distribution, network topology, and federation size, on the sustainability of DFL systems. Besides, a sustainability-aware aggregation algorithm (\textit{GreenDFL-SA}) and a node selection algorithm (\textit{GreenDFL-SN}) are developed to optimize energy efficiency and reduce carbon emissions in DFL training.\\
\textbf{Results:} Empirical experiments are conducted on multiple datasets, measuring energy consumption and carbon emissions at different phases of the DFL lifecycle. Results indicate that local training dominates energy consumption and carbon emissions, while communication has a relatively minor impact. Optimizing model complexity, using GPUs instead of CPUs, and strategically selecting participating nodes significantly improve sustainability. Additionally, using wired communication, particularly optical fiber, effectively reduces energy consumption during the communication phase, while integrating early stopping mechanisms further minimizes overall emissions.\\ 
\textbf{Conclusion:} The proposed \solution{} provides a comprehensive and practical approach for assessing the sustainability of DFL systems. Furthermore, it offers best practices for improving environmental efficiency in DFL, making sustainability considerations more actionable in real-world deployments.
\end{abstract}

\section{Introduction}
\label{sec:intro}


With the wide adoption of Artificial Intelligence (AI), especially the emergence of intelligent assistants based on Large Language Models (LLMs), Machine Learning (ML) has become deeply integrated into daily life~\cite{yao2024survey}. The \textit{neural scaling law} is no longer a prediction of real-world observations; it has come to be broadly recognized as a fundamental principle~\cite{kaplan2020scaling}. However, developing ever-larger ML models requires feeding them ever-increasing amounts of data. In the vanilla ML training process, data are typically centralized in a single entity or data center to facilitate training, thus, raw user data must be transferred and stored on a central server. This centralized approach, however, raises significant privacy concerns: users prefer not to expose their sensitive information~\cite{federatedlearning}. Meanwhile, legal and regulatory requirements increasingly restrict extensive data aggregation~\cite{beltran2023decentralized}. As a novel paradigm that mitigates these privacy challenges, Federated Learning (FL) has garnered substantial attention for its privacy-preserving capabilities and collaborative learning mechanisms~\cite{mcmahan2017fedavg}.

FL leverages a distributed training paradigm that distinguishes it from conventional ML training~\cite{federatedlearning}. In FL, data remain on local devices (clients), and each client trains a local model using its private data. These locally trained models, rather than the raw data, are then sent to a central server for aggregation and subsequent redistribution. However, such a Centralized FL (CFL) architecture suffers from the drawbacks of a single point of failure and potential bottlenecks at the central server. To overcome these limitations, Decentralized FL (DFL) removes the central server, employing peer-to-peer communication such that models are directly exchanged among nodes for aggregation~\cite{beltran2023decentralized}. By eliminating the client-server distinction, DFL mitigates the single-point-of-failure risk and offers greater system robustness. In DFL, data remain on each node for local training; afterward, models are exchanged and aggregated among neighboring nodes, proceeding recursively until the federated model converges. This paradigm not only addresses the server bottleneck but also enables more flexible network topologies and improved scalability~\cite{beltran2023decentralized}. However, FL is not immune to privacy and poisoning risks, since model updates can still leak sensitive information through inference or reconstruction attacks~\cite{feng2024dart}. In practice, CFL has been adopted in applications such as Google Gboard~\cite{hard2019federatedlearningmobilekeyboard} and Apple Siri~\cite{granqvist2020improvingondevicespeakerverification}, while DFL remains largely at the research stage.

As a specialized class of software systems, AI systems are increasingly becoming a critical topic in sustainability research~\cite{lacoste2019quantifyingcarbonemissionsmachine}. The scaling laws that drive larger models inherently imply massive computational demands. Supporting these large-scale computations consumes substantial amounts of energy and has significant environmental impacts, particularly greenhouse gas emissions~\cite{ding2024sustainable}. However, current studies on the sustainability of AI software systems have predominantly concentrated on energy consumption in traditional centralized ML paradigms, leaving a gap in rigorous frameworks for assessing and quantifying sustainability in distributed and especially fully decentralized FL systems~\cite{lynn2023}.

In DFL systems, nodes may be geographically distributed across different regions or even countries, necessitating an independent evaluation of each node's energy consumption and carbon emissions. Such complexity is not commonly encountered in centralized AI systems, where training occurs in a single data center. Moreover, in DFL, each node is responsible not only for local model training but also for model comm and aggregation. Research on centralized ML typically focuses on the energy consumption and environmental impacts of model training alone, largely overlooking the substantial energy costs of communication and model aggregation. While studies have explored sustainability in CFL by measuring carbon emissions, research on DFL remains limited~\cite{lynn2023}. In particular, there is a lack of comprehensive solutions for systematically measuring sustainability aspects in DFL and integrating energy-efficient strategies into node selection and aggregation processes. Furthermore, heterogeneity presents additional hurdles in DFL. Nodes may vary regarding local data distributions, tasks, security requirements, model architectures, and hardware capabilities. This multi-layered heterogeneity makes it challenging to develop standardized methodologies for evaluating and reducing the environmental footprint of DFL systems, further highlighting the need for targeted research on sustainability-aware DFL frameworks.

To address the current research gap in assessing the sustainability of DFL systems, this paper proposes the \solution{} framework for quantitatively analyzing and evaluating DFL energy consumption and environmental impacts. The framework takes account of the entire lifecycle of the DFL model training process, encompassing local training,  communication, and model aggregation. The main contributions are as follows:
\begin{itemize}
    \item \textbf{Quantitative Sustainability Framework:} This paper proposes an operational framework, named \solution{}, for comprehensively computing energy consumption and equivalent CO$_2$ emissions in DFL systems, enabling a quantitative assessment of their sustainability and environmental impacts. A prototype of the proposed framework is implemented and integrated into an open-source DFL system, \textit{Nebula} \footnote{Code available at: https://github.com/CyberDataLab/nebula}, demonstrating its feasibility and effectiveness in real-world scenarios.
    \item \textbf{Sustainability-Aware Aggregation and Node Selection Algorithm:} A sustainability-aware aggregation algorithm (\textit{GreenDFL-SA}) is developed to optimize the environmental impact of the aggregation process, ensuring a more energy-efficient model update. Additionally, a node selection algorithm (\textit{GreenDFL-SN}) is introduced to determine participating nodes during each training round dynamically. This method utilizes a voting mechanism, allowing nodes to decide which participants continue training based on their reported sustainability metrics, thereby reducing overall energy consumption while maintaining model performance.
    \item \textbf{Empirical Analysis:} Through extensive experiments and case studies, the paper applies the proposed framework to identify and analyze factors affecting the sustainability of DFL, offering practical insights into energy consumption trade-offs and carbon footprint reduction strategies.    
    \item \textbf{Best Practices for Sustainable DFL:} This paper provides recommendations for enhancing the sustainability of DFL systems, including strategies for optimizing model training strategy and leveraging renewable energy sources.

\end{itemize}

The remainder of this paper is organized as follows. Section~\ref{sec:related} contains findings from the literature review on sustainability in FL. Section~\ref{sec:design} introduces the proposed \solution{} framework. Section~\ref{sec:implement} details its implementation. Section~\ref{sec:exp} presents the experimental results and analyzes key findings. Section~\ref{sec:discussion} discusses best practices for sustainable DFL. Threats to validity are discussed in Section~\ref{sec:threatstovalidity}.  Finally, Section~\ref{sec:conclusion} summarizes the conclusions and outlines directions for future research.

\begin{table*}[t]
\centering
\caption{Comparison of Related Work Regarding Energy Consumption , Environmental Impacts, and Software Aspects in ML, FL, and Software Systems}
\label{tab:related_work}  
\begin{tabular}{@{} p{1.3cm}p{1.5cm}p{1.6cm}p{3.9cm}p{4.0cm}p{3.2cm} @{}}
\hline 
\textbf{Work} & \textbf{Paradigm} & \textbf{Metrics} & \textbf{Software/Engineering Aspect} & \textbf{Highlights} & \textbf{Optimization Strategies} \\
\hline 
\cite{qiu2021federatedlearningsaveplanet} 2021 & ML, CFL & CO$_2$ & Simulation scripts only, no reusable software & Quantify carbon emission estimation of CFL. \newline Compare the carbon emission of CFL and ML.  & None \\
\hline 
\cite{qiu2023lookcarbonfootprintfederated} 2023 & CFL &  Energy, CO$_2$ & Real hardware experiments, methodology only (no framework) & Introduce a generalized methodology to compute the carbon footprint of CFL. &  None \\
\hline 
\cite{energyDFL} 2023 & ML, CFL, \newline DFL & Energy, CO$_2$ & Analytical model, no system implementation & Propose frameworks for the calculation of energy consumption and carbon emission in ML, CFL, and DFL. & None \\ 
\hline 
\cite{lynn2023} 2024 & CFL & Hardware efficiency, carbon intensity & Implementation in a CFL platform & Qualitative sustainability analysis of CFL & None \\    
\hline
\cite{thakur2025greenfl} 2025 & FL & Multi-dimensional & Survey only, no software/tool contribution & Taxonomy of Green FL methods & Guidelines for energy-efficient FL \\ 
\hline
\cite{lago2025} 2025 & Software systems & Sustainability indicators & Developed toolkit for software sustainability assessment & Toolkit with multi-dimensional indicators & Conceptual modeling \\ 
\hline
\cite{omar2024} 2024 & ML-enabled systems & Accuracy/ energy trade-off & Concept drift detection algorithm, not a full system & Trade-offs between monitoring accuracy and energy & Algorithm-level optimization \\ 
\hline
\cite{juiz2024} 2024 & Data centers & CiS2, ISO 30134-4 & KPI metrics for datacenters, no software & Standardized sustainability metrics & Resource scheduling \\ 
\hline
\cite{andringa2025} 2025 & Cloud-native systems & Energy profiling & Cloud-native software experiments & Energy profiling of software systems & Architectural optimization \\ 
\hline
\cite{domingo2025} 2025 & ML & Energy per epoch & Experimental training optimizations, no general software framework & Energy-efficient training (layer freezing, quantization) & Algorithm-level energy reduction \\ 
\hline
This work & CFL, DFL  &Energy (kWh), CO$_2$ & Implementation on Nebula DFL platform & Quantitative sustainability assessment of DFL. Development of sustainability-aware aggregation (GreenDFL-SA) and node selection (GreenDFL-SN) algorithms & Aggregation + node scheduling optimization \\    
\hline
\end{tabular}  
\end{table*}

\section{Related Work}
\label{sec:related}
This section provides a review of the literature concerning the energy consumption and environmental impacts associated with FL. \tablename~\ref{tab:related_work} summarizes the research findings on the environmental sustainability aspect of ML and FL systems. 

Sustainability in software engineering has been conceptualized as a multi-dimensional construct by the Karlskrona Manifesto for Sustainability Design~\cite{becker2015karlskrona}. It identifies environmental, social, economic, technical, and individual dimensions of sustainability. The environmental dimension emphasizes minimizing the negative ecological impact of software systems throughout their lifecycle. This includes the consumption of natural resources (such as energy), the generation of emissions (such as CO$_2$), and the long-term ecological consequences of operating large-scale computing infrastructures.

Although the sustainability of traditional ML has attracted attention in academia, research on the sustainability of FL remained relatively scant. \cite{patterson2022carbonfootprintmachinelearning} indicated that the geographic location of ML training servers, the composition of the energy grid, the duration of the training, and even the specific brand and hardware type significantly affected overall carbon emissions.  Algorithm-level techniques such as runtime layer freezing, quantization, and early stopping, as proposed by Domingo-Reguero et al.~\cite{domingo2025}, can reduce the training energy of ML models. Even though these works focused on ML, they inspired subsequent research on the sustainability of FL.

A pioneering effort in FL sustainability was presented in \cite{qiu2021federatedlearningsaveplanet}, which offered the first systematic investigation into the carbon footprint of centralized FL (CFL). This work introduced a model for quantifying the carbon footprint of CFL, thus enabling an in-depth examination of how different CFL design choices influenced carbon emissions. In addition, it compared CFL’s carbon footprint with that of centralized ML. Subsequent research generalized the carbon emissions calculation method across various CFL configurations and tested it on real CFL hardware setups, examining how different settings, model architectures, training strategies, and tasks affect sustainability \cite{qiu2023lookcarbonfootprintfederated}. Feng et al. \cite{lynn2023} expanded the trustworthiness framework for CFL by introducing sustainability as a new evaluation pillar, thereby addressing all seven key AI requirements outlined by the European Commission’s High-Level Expert Group on AI. In this expanded framework, sustainability was evaluated through qualitative metrics such as hardware efficiency, federation complexity, and the carbon intensity of local energy grids, offering insights into the environmental footprint of FL systems. However, this study employed a qualitative approach and did not provide a quantitative analysis of FL’s sustainability. Beyond these individual contributions, a survey by Thakur et al. \cite{thakur2025greenfl} synthesized the emerging field of Green FL. The survey identified energy- and carbon-aware techniques across the FL lifecycle, but also emphasized that most existing efforts remained fragmented and largely restricted to CFL, with limited attention to decentralized settings.

A further contribution introduced a framework for analyzing energy consumption and carbon emissions in ML, CFL, and DFL contexts \cite{energyDFL}. This work quantified both the energy consumption and the equivalent carbon emissions associated with classical FL approaches as well as consensus-based decentralized methods, pinpointing optimal thresholds and operational parameters that could make FL designs more environmentally friendly. This study proposed a general computational framework but did not differentiate between energy consumption and carbon emissions from training versus aggregation. Moreover, it assumed that each node's energy consumption was known, a condition that is often infeasible in practice. 

In parallel, the broader software and data center communities have developed sustainability assessment frameworks and standardized metrics. Lago et al.~\cite{lago2025} proposed a modeling toolkit that consolidates more than a decade of experience in sustainability assessment for software systems. Their framework emphasizes multi-dimensional indicators and provides guidelines for modeling trade-offs across environmental, economic, and technical dimensions. While this toolkit is versatile, it was designed primarily for centralized software systems and lacks mechanisms to capture the dynamic and distributed nature of DFL.Omar et al.~ \cite{omar2024} evaluated the sustainability from a monitoring perspective, analyzing the trade-offs between accuracy and energy consumption in concept drift detection for ML-enabled systems. Their work highlights the importance of balancing functional performance with energy efficiency, yet it remains confined to specific ML monitoring tasks and does not provide a generalized framework applicable to collaborative training. Juiz et al.~\cite{juiz2024} advanced the state of practice in sustainable data centers by defining ISO-compliant metrics such as the consolidated CiS2 metric. These metrics enable benchmarking of energy efficiency and carbon emissions at the datacenter scale, but they assume controlled environments with centralized resource management, which is in stark contrast to the heterogeneity and autonomy of edge devices in DFL. Similarly, Andringa et al.~\cite{andringa2025} investigated the energy consumption of cloud-native software, uncovering how architectural choices influence runtime energy usage. Their findings reinforce the role of software architecture in shaping sustainability outcomes but remain limited to cloud-native deployments.

Industry tools such as Google’s Carbon Footprint~\cite{google_carbon_footprint}, IBM's Cloud Carbon Calculator~\cite{ibm_cloud_carbon_calculator} and Salesforce's Net Zero Cloud~\cite{salesforce_net_zero_cloud} provide carbon accounting and resource optimization capabilities in cloud datacenters, but they are tailored for centralized cloud environments and offer limited support for edge computing and IoT scenarios.

In conclusion, existing research and tools on software sustainability primarily focused on centralized systems, with limited attention paid to DFL. Although DFL-focused works identified various factors affecting energy efficiency and carbon emissions, their proposed computational methods lacked practical operability. In addition, these studies often overlooked the renewable energy substitution rate in the nodes’ energy sources, relying instead on broad estimates of the local grid’s carbon intensity, which introduced inaccuracies. Moreover, existing work provided limited practical guidance for training real-world DFL systems, as it did not propose an algorithm that used sustainability metrics to optimize node selection in DFL.

\section{The \solution{} Framework}
\label{sec:design}
This section delves into the detailed methodologies of the \solution{} framework for calculating energy consumption and carbon emissions within DFL environments. 

\subsection{Research Methodology}
\label{sec:researchmethodology}
This section presents the research methodology, including the defined research scope, the formulated research questions, and the applied methodological approach.

\subsubsection{Research Scope}
This work investigates the environmental sustainability of DFL systems, focusing on energy consumption and carbon emissions. Other sustainability dimensions, such as social or economic factors, are beyond its scope. While the sustainability of a DFL system can be affected by various stages of the ML lifecycle, including data collection, model development, training, and deployment, this work concentrates on the learning stage, as it represents the dominant share of runtime energy usage in DFL. The developed models and prepared datasets are assumed to be available to all nodes before training.

\subsubsection{Research Questions}
To structure the investigation, this work formulates three research questions. These research questions address the main sources of environmental impact in DFL, the conditions under which sustainability varies, and the potential of optimization strategies.
This work is guided by the following research questions:
\begin{itemize}
    \item \textbf{\textit{RQ1}}: \textbf{Which phase of the DFL lifecycle, training, communication, or aggregation, contributes most to energy consumption and carbon emissions?} This question aims to identify the main source of environmental impact within the iterative process of DFL. Understanding the relative contribution of each phase provides guidance on where optimization efforts should be directed.
    \item \textbf{\textit{RQ2}}: \textbf{How do system- and environment-related factors, such as network topology, data distribution, model architecture, and energy carbon intensity, affect the sustainability of DFL?} This question examines the sensitivity of DFL sustainability to both design choices and external conditions. The results can reveal which parameters are most influential and should be considered when deploying DFL systems.
    \item \textbf{\textit{RQ3}}: \textbf{Can sustainability-aware aggregation and node selection strategies reduce the environmental footprint of DFL compared to standard approaches?} This question evaluates the effectiveness of algorithmic modifications that incorporate sustainability metrics into decision-making. It tests whether such strategies can achieve measurable reductions in energy consumption and carbon emissions.
\end{itemize}

\subsubsection{Methodological Process}
To address these research questions, the work follows a five-step methodological process. First, a literature review is conducted to examine sustainability in ML and FL, together with established frameworks for software systems, data centers, and cloud-native applications. Secondly, building on these insights, the \solution{} framework is designed, defining relevant metrics and modeling the DFL lifecycle in terms of training, communication, and aggregation. Based on this framework, sustainability-aware aggregation (\textit{GreenDFL-SA}) and node selection (\textit{GreenDFL-SN}) algorithms are developed that incorporate environmental metrics into decision-making. The framework and algorithms are then implemented within the \textit{Nebula} DFL platform to ensure practical applicability. Finally, a series of experiments across datasets, network topologies, and geographic settings are carried out to evaluate the framework and to answer RQ1 to RQ3.

\begin{figure*}[t]
    \centering
    \includegraphics[width=\linewidth]{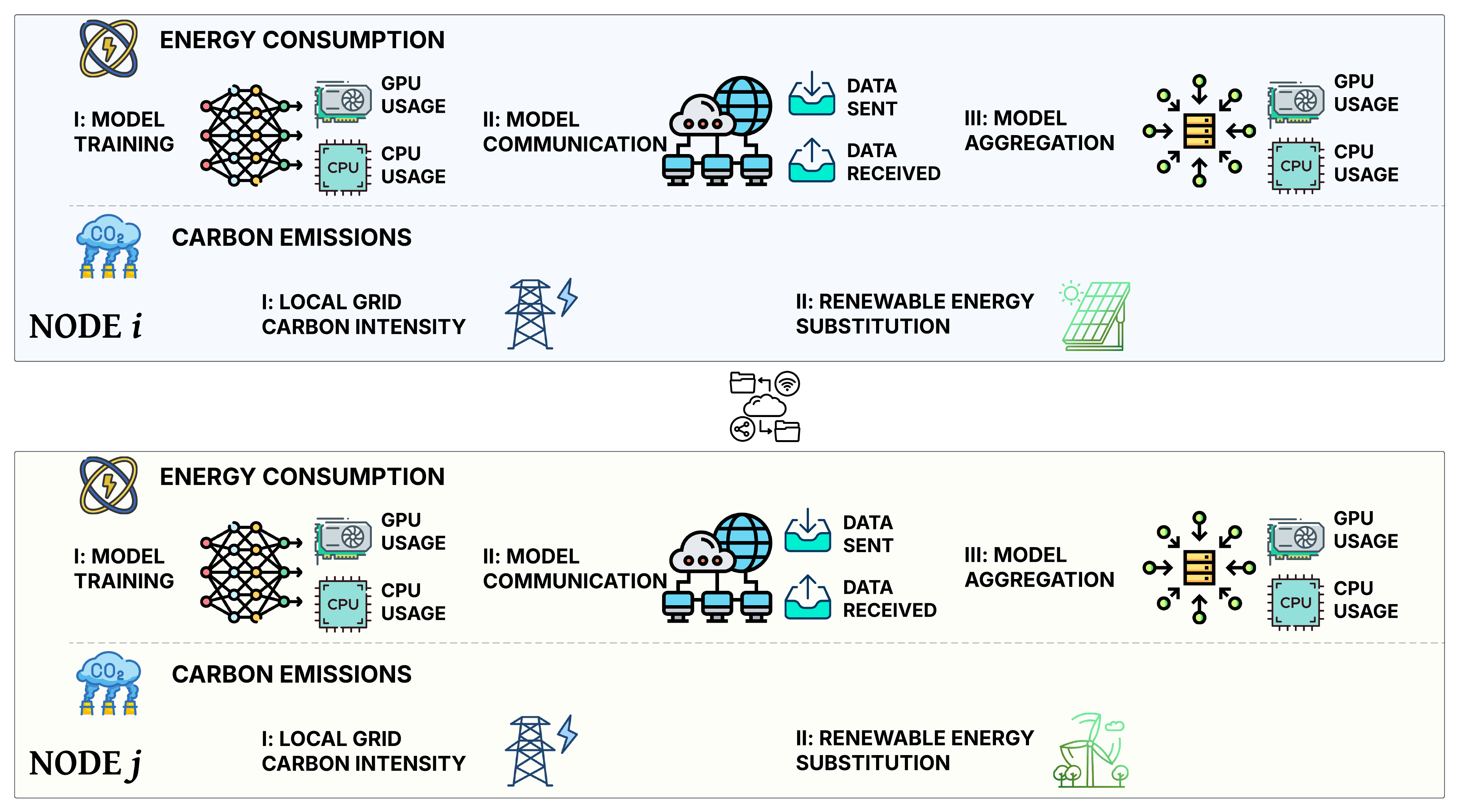}
    \caption{Overall Architecture of the \solution{} Framework}
    \label{fig:arch}
\end{figure*}

\subsection{\solution{} Overview}

The \solution{} framework models the lifecycle of DFL in three stages: training, communication, and aggregation. These stages describe the main processes in most FL systems~\cite{beltran2023decentralized}.  Variants such as fog computing, edge hierarchies, or grid-based federations can be mapped to these processes.

\begin{itemize}
    \item \textbf{Model Training}: This phase involves the local computation by DFL nodes, where models are trained on distributed nodes using local data. This is the most computationally intensive phase, often resulting in significant energy use and associated carbon emissions.

     \item \textbf{Communication}: Communication refers to exchanging local models among nodes in DFL systems. This stage involves sending the local model to other nodes and receiving models from them. Since the primary energy consumption arises from network communication, the size of the model plays a critical role.

    \item \textbf{Model Aggregation}: When a node receives the desired models from other nodes, it uses an aggregation algorithm, commonly FedAvg, to aggregate its local model with the received models. This aggregation's computational load depends on the number of models being merged and the size of their parameters. Consequently, the scale and topology of the DFL system often play a critical role in determining energy consumption at this stage.
\end{itemize}

In line with the DFL learning lifecycle, as shown in \figurename~\ref{fig:arch}, \solution{} divides the sustainability analysis of DFL systems into three phases: model training, communication, and model aggregation. Each phase is examined for its energy consumption and carbon emissions, offering a holistic view of the environmental impact of DFL systems. The following subsections decompose and explain each phase in \solution{} from energy consumption and carbon emissions perspectives, providing a practical computational framework.

\subsection{Calculation of Energy Consumption}
\label{sec:Calculation of Energy Consumption}

The energy consumption of \solution{} can be broken down into three components: model training, model communication, and model aggregation. For the training and aggregation stages, \solution{} adopts a quantifiable and implementable approach by primarily considering each node’s hardware architectures (\ie device models and different accelerators), hardware resource utilization, and computation time. For model communication, \solution{} focuses on the volume of data being sent and received, as well as the energy required to transfer one byte of data. Thus, the total energy consumption could be modeled as:
\begin{equation}
\label{eq:enenrgy_total}
E = E^{(T)} + E^{(C)} + E^{(A)}
\end{equation}
where $E^{(T)}, E^{(C)}, E^{(A)}$ are the energy consumption in the model training, communication, and aggregation, respectively. While studies, such as \cite{domingo2025}, employ normalized metrics (e.g., energy per epoch or FLOPs per Watt) to assess efficiency, these primarily capture computational operations and do not account for the communication stage, which is an important component in DFL. Therefore, this work adopts absolute energy consumption and carbon emissions as the main sustainability indicators.

Before presenting the detailed energy consumption model, \tablename~\ref{tab:notations} summarizes the key notations used in the subsequent equations.

\begin{table}[h]
    \centering
    \caption{Table of Notations}
    \label{tab:notations}
    \begin{tabular}{@{} p{1cm}p{7cm} @{}}
        \toprule
        \textbf{Symbol} & \textbf{Definition} \\
        \midrule
        \( K \) & Number of nodes in the system \\
        \( n \) & Number of global learning rounds \\
        \( PUE_k \) & Power Usage Effectiveness of node \( k \), measuring power efficiency \\
        \( TDP_k \) & Thermal Design Power of node \( k \), indicating peak heat dissipation \\
        \( \beta_{k,t}^{(T)} \) & CPU utilization rate during training at node \( k \) in round \( t \) \\
        \( T_{k,t}^{(T)} \) & Duration of local training at node \( k \) in round \( t \)\\
        \( P_t^{(\text{GPU})} \) & Power consumption of GPU in round \( t \)\\
        \( \beta_{k,t}^{(A)} \) & CPU utilization rate during aggregation at node \( k \) in round \( t \) \\
        \( T_{k,t}^{(A)} \) & Duration of model aggregation at node \( k \) in round \( t \) \\
        \( B_{k,t}^{\text{sent}} \) & Total bytes sent by node \( k \) in round \( t \) \\
        \( B_{k,t}^{\text{recv}} \) & Total bytes received by node \( k \) in round \( t \) \\
        \( E_{\text{byte},k}^{(C)} \) & Energy consumption per byte transmitted at node \( k \)\\
        \bottomrule
    \end{tabular}
\end{table}

\subsubsection{Model Training Phase Energy Consumption}
The calculation of energy consumption for local model training in a node depends on the type of accelerator utilized, such as CPUs or GPUs. During the training process, nodes may employ only CPUs or a combination of CPUs and GPUs. The total energy consumption for local training is determined by summing the energy consumed by both components, with GPU energy consumption considered zero when GPUs are not in use. The energy consumption for CPU-based and GPU-based training is formulated in Equation \eqref{eq:energy_training}. 

\begin{equation}
\label{eq:energy_training}
\begin{aligned}
    E^{(T)} &= E_{\text{CPU}}^{(T)} + E_{\text{GPU}}^{(T)} \\
    E_{\text{CPU}}^{(T)} &= \sum_{k=1}^{K} \sum_{t=1}^{n} PUE_k \cdot TDP_k \cdot \beta_{k,t}^{(T)} \cdot T_{k,t}^{(T)} \\
    E_{\text{GPU}}^{(T)} &= \sum_{k=1}^{K} \sum_{t=1}^{n} P_t^{(\text{GPU})} \cdot T_{k,t}^{(T)}
\end{aligned}
\end{equation}

Equation~\ref{eq:energy_training} quantifies the total CPU energy consumption during local training. The calculation considers the power usage effectiveness (\( PUE_k \)), the thermal design power (\( TDP_k \)), CPU utilization rate (\( \beta_{k,t}^{(T)} \)), and the training duration (\( T_{k,t}^{(T)} \)). Besides, it calculates the total GPU energy consumption across all nodes and rounds, considering the training time (\( T_{k,t}^{(T)} \)) and GPU power consumption (\( P_t^{(\text{GPU})} \)). Rather than relying on the GPU’s TDP, which denotes a vendor-specified upper bound and typically overestimates actual runtime consumption, this work derives energy estimates from measured GPU power values sampled during training (e.g., via \texttt{nvidia-smi}), ensuring a more accurate representation of real energy usage.

\subsubsection{Communication Phase Energy Consumption}
In DFL systems, energy consumption is not limited to computation but also arises from communication. During model updates, nodes exchange data, contributing to overall energy consumption. The energy consumption for communication is formulated as in Equation \eqref{eq:communication_energy}:

\begin{equation}
\label{eq:communication_energy}
E^{(C)} = \sum_{k=1}^{K} \sum_{t=1}^{n} \left[ \left(B_{k,t}^{\text{sent}} + B_{k,t}^{\text{recv}}\right) \cdot E_{\text{byte},k}^{(C)} \right]
\end{equation}

Equation \eqref{eq:communication_energy} accounts for both sent (\( B_{k,t}^{\text{sent}} \)) and received (\( B_{k,t}^{\text{recv}} \)) data at each node during each training round. The energy per byte transferred (\( E_{\text{byte}}^{(C)} \)) is multiplied by the total data exchanged.

\subsubsection{Model Aggregation Phase Energy Consumption}
The energy consumption during the aggregation phase is computed as follows:  

\begin{equation}
\label{eq:aggregation_energy}
\begin{aligned}
    E^{(A)} &= E_{\text{CPU}}^{(A)} + E_{\text{GPU}}^{(A)} \\
    E_{\text{CPU}}^{(A)} &= \sum_{k=1}^{K} \sum_{t=1}^{n} PUE_k \cdot TDP_k \cdot \beta_{k,t}^{(A)} \cdot T_{k,t}^{(A)} \\
    E_{\text{GPU}}^{(A)} &= \sum_{k=1}^{K} \sum_{t=1}^{n} P_t^{(\text{GPU})} \cdot T_{k,t}^{(A)}
\end{aligned}
\end{equation}

Equation~\ref{eq:aggregation_energy} follows a similar methodology to the energy consumption calculation during training, incorporating both CPU and GPU energy consumption depending on the type of accelerator utilized.

By integrating these three aspects, \solution{} offers a comprehensive and operational framework for evaluating the sustainability of DFL systems.

\subsection{Calculation of Carbon Emissions}
After computing the energy consumption of the DFL system, the next step is to estimate its carbon emissions. This process consists of two key steps. Firstly, the carbon intensity (\( CI \)) of the energy used by each node must be identified, as it varies depending on the regional energy grid carbon intensity and the proportion of renewable energy utilized. Secondly, the total carbon emissions of the DFL system are obtained by multiplying the energy consumption of each node by its corresponding carbon intensity. The total carbon emissions of the DFL system are computed as follows:

\begin{equation}
\label{eq:carbon_emission}
C = \sum_{k=1}^{K} CI_k \cdot E_k
\end{equation}

where \( C \) represents the total carbon emissions (gCO$_2$) of the DFL system, \( CI_k \) denotes the carbon intensity (gCO$_2$/kWh) at node \( k \), and \( E_k \) represents the total energy consumption (kWh) of node \( k \). This equation provides a comprehensive assessment of the carbon footprint of a DFL system by aggregating emissions across all nodes.

Before presenting the carbon intensity and emissions calculations, \tablename~\ref{tab:cab_notations} summarizes the key notations used in the following equations.

\begin{table}[h]
    \centering
    \caption{Table of Notations}
    \label{tab:cab_notations}
    \begin{tabular}{@{} p{1.3cm}p{6.7cm} @{}}
        \toprule
        \textbf{Symbol} & \textbf{Definition} \\
        \midrule
        \( CI_k \) & Carbon intensity (gCO$_2$/kWh) at node \( k \) \\
        \( CI_{k, \text{local}} \) & Local carbon intensity (gCO$_2$/kWh) at node \( k \) \\
        \( CI_{k, \text{renewable}} \) & Carbon intensity for renewable energy, approximated as zero \\
        \( R_k \) & Renewable energy ratio at node \( k \) \\
        \bottomrule
    \end{tabular}
\end{table}

\subsubsection{Carbon Intensity}

Carbon intensity is typically defined as the ratio of CO$_2$ emissions to energy consumption (gCO$_2$/kWh). Calculating carbon intensity is crucial because it quantifies the CO$_2$ emissions produced per unit of energy consumed. Carbon intensity varies according to the geographic location of the nodes, as different regions may depend on various energy sources with distinct carbon footprints. The energy grid used by a node can be determined by its location, typically defined by the latitude and longitude, aligning with the energy mix of the country where the node operates.

Besides, carbon intensity also depends on the renewable energy ratio of the nodes' local places. In this work, $CI_{k,\text{local}}$ is treated as an external parameter that characterizes the local grid mix at node $k$, obtained from official reports or databases. An increasing number of data centers and even households are integrating self-generated renewable energy as an alternative to the traditional power grid. Considering this factor allows for a more accurate assessment of carbon emissions. The higher the renewable energy ratio, the lower the carbon intensity of the region. The carbon intensity at a given node is computed as follows:

\begin{equation}
\label{eq:CI}
CI_k = CI_{k, \text{local}} \cdot (1 - R_k)
\end{equation}

\begin{equation}
\label{eq:CI_renewable}
CI_{k, \text{renewable}} \approx 0
\end{equation}

Equation~\ref{eq:CI} adjusts the local carbon intensity (\( CI_{k, \text{local}} \)) by the proportion of energy derived from the local grid (\( 1 - R_k \)). A higher renewable energy ratio (\( R_k \)) results in lower carbon intensity. This formulation does not redefine carbon intensity; instead, it uses the standard carbon intensity factor as a multiplier to translate energy consumption into total emissions, while accounting for the fraction of renewable energy. 
Equation~\ref{eq:CI_renewable} approximates the carbon intensity of renewable energy sources as zero, referring to the operational phase where emissions are negligible. It is acknowledged that lifecycle emissions exist due to infrastructure development (e.g., production of solar panels or wind turbines), yet these are comparatively minor and typically amortized over the long-term energy output.

\subsubsection{Carbon Emissions}
With the energy consumption and carbon intensity established, the total carbon emissions at each node are determined using the following equations. These equations integrate energy consumption for training, aggregation, and communication with the specific carbon intensity at each node.

\begin{equation}
\label{eq:carbon_emissions}
\begin{split}
C &= \sum_{k=1}^{K} CI_k \cdot E_k \\
  &= \sum_{k=1}^{K} CI_k \cdot E_k^{(T)} + E_k^{(C)} + E_k^{(A)}\\
  &= \sum_{k=1}^{K} \sum_{t=1}^{n} CI_k \cdot [PUE_k \cdot TDP_k \cdot (\beta_{k,t}^{(T)} \cdot T_{k,t}^{(T)} + \beta_{k,t}^{(A)} \cdot T_{k,t}^{(A)}) \\
  & + P_t^{(\text{GPU})} \cdot (T_{k,t}^{(T)} +  T_{k,t}^{(A)} ) + (B_{k,t}^{\text{sent}} + B_{k,t}^{\text{recv}}) \cdot E_{\text{byte},k}^{(C)}]
\end{split}
\end{equation}

As shown in Equation ~\ref{eq:carbon_emissions}, by incorporating carbon intensity and energy consumption, \solution{} provides a comprehensive assessment of the carbon emissions of DFL systems, offering insights for optimizing energy efficiency and reducing environmental impact.

\subsection{Sustainability-Aware Aggregation (\textit{SA})}
The above analysis serves as the theoretical foundation, providing a cornerstone for exploring sustainability-oriented optimizations in DFL. Building on this basis, two optimization directions are proposed, focusing on model aggregation and node selection.

To optimize the environmental impact of the aggregation process in DFL, this paper proposes a sustainability-aware aggregation algorithm, \textit{GreenDFL-SA}, as shown in Algorithm~\ref{alg:carbon_aware_aggregation}.

In a DFL system, nodes exchange model updates and carbon emission values with their neighboring nodes. Algorithm~\ref{alg:carbon_aware_aggregation} allows each node to dynamically select a subset of its neighbors for model aggregation based on their carbon emissions. By filtering out high-emission nodes, the system promotes sustainable collaboration, effectively reducing the overall carbon footprint of model training.

\begin{algorithm}[H]
\caption{\textit{GreenDFL-SA} Algorithm}
\label{alg:carbon_aware_aggregation}
    \begin{algorithmic}[1]
        \Require Total nodes \( K \), Current node \( i \), Neighbor set \( \mathcal{N}_i \), Carbon threshold \( C_{\text{thresh}} \)
        \State \textbf{Initialize} selected neighbor set \( S_i \gets \emptyset \)
        \State \textbf{Broadcast} local model \( M_i^{(t)} \) and carbon emission \( C_i^{(t)} \) to neighbors \( \mathcal{N}_i \)
        \State \textbf{Receive} models \( \{M_j^{(t)}\}_{j \in \mathcal{N}_i} \) and emissions \( \{C_j^{(t)}\}_{j \in \mathcal{N}_i} \)
        \For{each neighbor \( j \in \mathcal{N}_i \)}
            \If{ \( C_j^{(t)} \leq C_{\text{thresh}} \) } 
                \State \( S_i \gets S_i \cup \{ j \} \) \Comment{Select neighbor \( j \) for aggregation}
            \EndIf
        \EndFor
        \State \textbf{Compute aggregated model:}
        \begin{equation}
            M_i^{(t+1)} = \sum_{j \in S_i} w_j M_j^{(t)}
        \end{equation}
        \State \textbf{Update} local model \( M_i^{(t+1)} \)
    \end{algorithmic}
\end{algorithm}

At the start of each training round, each node broadcasts its local model and carbon emissions to its neighbors and receives the same information. Each node then evaluates the reported emissions of its neighbors against a threshold that is determined during an initialization phase based on the distribution of observed energy consumptions (e.g., set to the 75th percentile). This threshold is predefined and can be configured by the user during the setup phase. Only models with emissions below this threshold are included in the aggregation process.

All decisions are made locally by each node without centralized coordination. Nodes that are not selected for aggregation still proceed with the next round of training. Therefore, this algorithm impacts only the sustainability of the aggregation phase. The selected neighbors’ models are then weighted and aggregated to update the local model. This adaptive selection strategy ensures that nodes with lower environmental impact have a greater influence on the global model, fostering a more energy-efficient and sustainable DFL process.

\subsection{Sustainability-Aware Node Selection (\textit{SN})}
The \textit{GreenDFL-SA} optimizes energy consumption during the aggregation phase. However, the training phase is the most computationally intensive stage in the DFL learning lifecycle, making it a critical target for energy optimization. To address this, this work proposes the Sustainability-Aware Node Selection Algorithm, called \textit{GreenDFL-SN}, shown in Algorithm~\ref{alg:sustainability_node_selection}, which aims to reduce energy consumption during local training by selectively enabling only the most sustainable nodes to participate in each training round.

\begin{algorithm}[b]
    \caption{\textit{GreenDFL-SN} Algorithm}
    \label{alg:sustainability_node_selection}
    \begin{algorithmic}[1]
        \Require Number of nodes \( K \), Carbon Intensity (CI) reports from all nodes
        \Ensure Set of selected nodes for next training round \( S \)
        
        \State \textbf{Initialize} vote counters $v_i \gets 0$ for all $i \in \{1,\dots,K\}$, and $S \gets \emptyset$
        \State \textbf{Broadcast:} each node $i$ sends $CI_i$ to all neighbors $j \in \mathcal{N}_i$
        \State \textbf{Neighbor voting:} \For{each node $i \in \{1,\dots,K\}$}
            \For{each neighbor $j \in \mathcal{N}_i$}
                \If{$CI_j \le CI_i$}
                    \State $v_j \gets v_j + 1$ \Comment{$i$ votes positively if $j$ is at least as efficient}
                \EndIf
            \EndFor
        \EndFor

        \State \textbf{Majority decision:} \For{each node $i \in \{1,\dots,K\}$}
            \If{$v_i >= \frac{|\mathcal{N}_i|}{2}$} \Comment{at least 50\% of neighbors voted}
                \State $S \gets S \cup \{i\}$
            \Else
                \State \textbf{exclude} $i$ from next round
            \EndIf
        \EndFor

        \State \Return $S$
    \end{algorithmic}
\end{algorithm}

The \textit{GreenDFL-SN} algorithm operates independently from the sustainability-aware aggregation (\textit{GreenDFL-SA}) and focuses on selecting participants for each training round. The \textit{GreenDFL-SN} algorithm ensures that nodes with higher carbon efficiency are prioritized after each training round for continued participation. At the end of a training round, each node reports its  carbon intensity, representing the environmental impact of its energy consumption. Based on these reports, nodes collectively decide which peers should continue in the next round. The decision is made in a distributed manner: each node compares the reported carbon intensity of its neighbors with its own. If a neighbor’s intensity is higher, it casts a negative vote; otherwise, it casts a positive vote. A node is retained if it receives positive votes from at least half of its neighbors; otherwise, it is excluded from the subsequent round.

Assume a 5-node line topology $A\!-\!B\!-\!C\!-\!D\!-\!E$ with neighbor sets 
$\mathcal{N}_A=\{B\}$, $\mathcal{N}_B=\{A,C\}$, $\mathcal{N}_C=\{B,D\}$,  
$\mathcal{N}_D=\{C,E\}$, $\mathcal{N}_E=\{D\}$.  
Reported carbon intensities are $CI_A=150$, $CI_B=180$, $CI_C=220$,  
$CI_D=260$, $CI_E=140$ (gCO$_2$/kWh).  Each node votes \emph{positively} for a neighbor $j$ if $CI_j \le CI_i$, otherwise casts a \emph{negative} vote.  
A node is retained if it receives positive votes from \emph{at least half} of its neighbors, otherwise it is excluded. 

The votes are as follows:  

$B \!\to\! A$: positive, $A \!\to\! B$: negative;  

$C \!\to\! B$: positive, $B \!\to\! C$: negative; 

$D \!\to\! C$: positive, $C \!\to\! D$: negative;  

$E \!\to\! D$: negative, $D \!\to\! E$: positive.  

The final voting results are: $A:1/1$ (retain), $B:1/2$ (retain), $C:1/2$ (retain),  $D:0/2$ (exclude), $E:1/1$ (retain). Hence, the next-round training set is $\{A,B,C,E\}$. D is excluded from the training yet retained as a bridge node for model relaying.

This voting process allows the system to dynamically filter out high-carbon nodes, reducing the overall carbon footprint of the DFL system. By iteratively applying this voting-based selection, \textit{GreenDFL-SN} promotes energy-efficient participation in DFL without requiring central coordination.

\section{Framework Implementation}
\label{sec:implement}
\solution{} provides a comprehensive and operational framework for assessing the environmental sustainability of DFL systems. This section details its implementation and integration into the \textit{Nebula} DFL platform, including parameters acquisition and metrics computation.

\begin{figure*}[t]
    \centering
    \includegraphics[width=\linewidth]{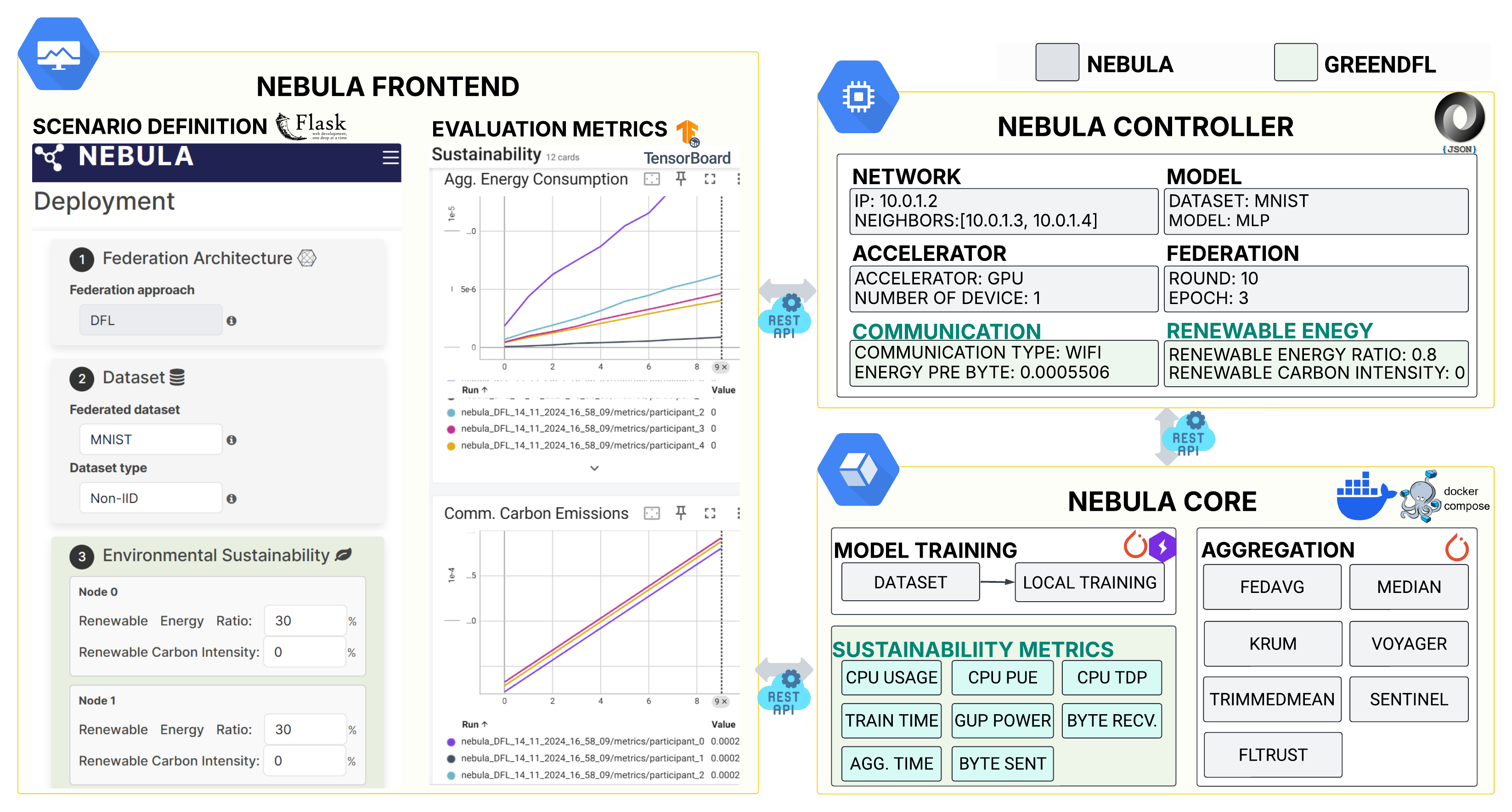}
    \caption{Integration of \solution{} Framework to the \textit{Nebula} Platform}
    \label{fig:nebula_architecture}
\end{figure*}

\begin{figure}[h!]
    \centering
    \includegraphics[width=\linewidth]{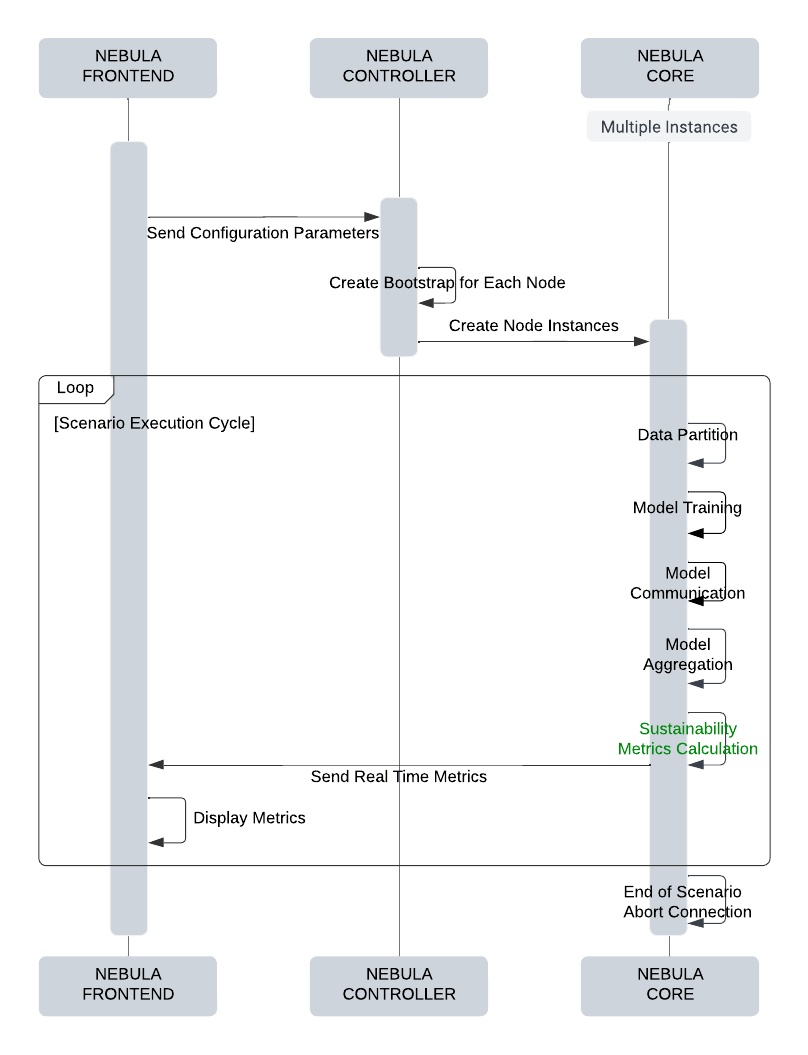}
    \caption{Sequence Diagram Showing the Interaction of \textit{Nebula} Components (Based on~\cite{fedstellar:2024})}
    \label{fig:nebula_Sequence}
\end{figure}

\subsection{\textit{Nebula} Platform}
\textit{Nebula}~\cite{fedstellar:2024} was chosen as the infrastructure for integrating \solution{} due to its flexibility and advanced functionality in deploying DFL systems. Existing FL software platforms mainly support the CFL paradigm. Although platforms, such as Flower~\cite{beutel2020flower}, also support DFL model training, their flexibility in topology and other aspects is not as strong as Nebula. Therefore, Nebula is chosen as the implementation platform. However, the computational methods presented in this work can, in principle, also be implemented on other platforms.

\textit{Nebula} is a versatile FL platform that supports multiple FL paradigms, allowing users to deploy various types of FL models, including CFL, semi-DFL, and DFL. Additionally, \textit{Nebula} offers multiple datasets and diverse model architectures for FL training. It enables users to customize the connectivity topology among nodes, enhancing its adaptability to different research and deployment needs. \textit{Nebula} consists of three components:  Frontend, Controller, and Core. Each component plays a distinct role in facilitating FL model deployment, configuration, and execution, enabling seamless integration with \solution{}. \figurename~\ref{fig:nebula_architecture} provides an overview of the three core modules of \textit{Nebula} and illustrates how \solution{} is implemented and integrated within these modules.

\begin{itemize}
    \item \textbf{Frontend}: Provides a user-friendly interface for configuring and deploying FL models. Users can specify various configurations such as FL architecture (CFL or DFL), dataset selection, data partitioning strategies, model architectures, communication topology, hardware accelerators, and aggregation algorithms. Besides, the frontend provides monitoring and visualization of the FL process. Users can track the real-time status of FL model training directly through the frontend interface.
    \item \textbf{Controller}: The controller acts as a middleware, translating user-configured scenarios into bootstrap configurations for individual nodes. It ensures that each node is correctly initialized with the designated configurations.
    \item \textbf{Core}: It is a fundamental component of the \textit{Nebula} platform, deployed into physical or virtualized devices by the controller. This component handles the execution of model training, communication, and aggregation based on the bootstrap configurations. It is responsible for orchestrating FL workflows across participats.
\end{itemize}

As shown in~\figurename~\ref{fig:nebula_Sequence}, the deployment process begins with the configuration of the scenario. Once the frontend submits the configuration, the controller instantiates a core module for each node and transfers the predefined parameters to these instances. The core is responsible for executing the local workflow, including data partitioning, model training, communication, aggregation, and sustainability metrics calculation.

\subsection{Implementation in the Frontend}
The frontend of \textit{Nebula} is built using the Flask framework~\cite{flask}, providing a web-based interface for configuring various FL parameters, including the DFL architecture, dataset selection, and training options. These configurations are transmitted to the controller via REST API, where they serve as initialization parameters for node bootstrapping.

To accommodate sustainability computations, the frontend implementation of \solution{} introduces the following additional configurations:
\begin{itemize}
 \item Communication Mode Selection: Users can specify whether wired or wireless communication is used. Additionally, the system allows users to define the energy consumption per byte transferred, corresponding to parameter \( E_{\text{byte},k}^{(C)} \)  in the Equations~\eqref{eq:communication_energy}.
 \item Local Renewable Energy Utilization: Users can input the proportion of energy sourced from self-produced renewable energy. This factor influences the carbon intensity calculation, corresponding to parameters (\( CI_{k, \text{local}} \)) and (\( CI_{k, \text{renewable}} \))
in Equations~\eqref{eq:CI} and~\eqref{eq:CI_renewable}.
\end{itemize}
Users only need to provide these three parameters. All other required parameters for \solution{} are automatically retrieved from the backend or determined at runtime. The current model adopts a static carbon intensity factor based on the national electricity grid mix, without considering short-term fluctuations such as renewable availability at specific times of day (e.g., solar panels generating only during daylight). While this simplification neglects temporal variations in renewable supply, it provides a stable and comparable baseline for evaluating the sustainability of DFL systems across regions.

The federation status and model performance are also visualized through the frontend. By integrating TensorBoard~\cite{tensorboard}, users can monitor the real-time execution of FL, including device utilization, as well as the training, validation, and testing performance of the model. Correspondingly, \solution{} utilizes REST API to receive sustainability metrics from the backend, including the energy consumption and carbon emissions associated with training, communication, and aggregation at each node.

\subsection{Implementation in the Controller}
The controller acts as a middleware component, bridging the frontend configurations with the FL nodes by converting user-defined parameters into structured initialization settings. It ensures that the FL system is correctly configured before execution. 

In the integration of \solution{}, the controller is responsible for handling sustainability-related parameters and incorporating them into the FL workflow. When the frontend transmits configurations via REST API, including communication mode and renewable energy utilization, the controller processes these inputs and encodes them into predefined JSON fields. These newly formatted fields are then injected into the node configuration files, ensuring that each node is bootstrapped with sustainability-aware parameters. This allows \solution{} to track and compute energy consumption and carbon emissions.

\subsection{Implementation in the Core}
The core module is responsible for executing of the DFL training process, including model training, aggregation, and communication. Thus, it plays a crucial role in computing the energy consumption and carbon emissions across the three phases of \solution{}.

\subsubsection{Energy Consumption and Carbon Emissions in the Training and Aggregation Phases}
To compute the energy consumption during training as described in Equation~\ref{eq:energy_training}, several key parameters must be automatically retrieved during the training process, including \( PUE_k \), \( TDP_k \), GPU power consumption (\( P_t^{(\text{GPU})} \)), CPU utilization (\( \beta_{k,t}^{(T)} \)), and training duration (\( T_{k,t}^{(T)} \)). CPU and GPU utilization are monitored at the system level using standard profiling tools, where the reported values indicate the percentage of total device capacity in use. Since FL is often deployed in edge computing scenarios where nodes are primarily dedicated to local training and typically run without other significant workloads, this system-level approximation serves as a practical measure of utilization.

The CPU model of each node must first be identified to obtain the \( TDP_k \) parameter. In this work, the Python platform library~\cite{python_platform} is utilized to retrieve the CPU model of each node automatically. The CPU model is then matched in a precompiled database~\cite{tdp} to obtain the corresponding \( TDP_k \) value. To measure CPU utilization related to model training, the psutil library~\cite{psutil} is employed to retrieve both CPU usage (\( \beta_{k,t}^{(T)} \)) and \( PUE_k \). If a model is trained using a GPU accelerator, the pynvml library~\cite{pynvml} is used to monitor GPU power consumption in real time. Additionally, the training duration (\( T_{k,t}^{(T)} \)) is recorded for each node, tracking the time interval from the start to the completion of local model training in each round of local training.  

Based on these parameters, \solution{} calculates the energy consumption for each node in each training round. Using Equation~\ref{eq:carbon_emissions}, the corresponding carbon emissions per training round are determined. 

The energy consumption of aggregation is computed analogously to training, 
with aggregation duration (\(T_{k,t}^{(A)}\)). Per-round values are accumulated across all nodes, and the totals are used to derive the overall carbon emissions from training and aggregation in the federated system.

\subsubsection{Energy Consumption and Carbon Emissions in the Communication Phase}
As shown in Equation~\ref{eq:communication_energy}, the energy consumption during the communication phase primarily depends on the amount of data sent and received by each node in every aggregation round, as well as the energy consumption per byte transmitted.

To obtain these values, the psutil library is used to retrieve the data communication metrics of each node during each round $(B_{k,t}^{\text{sent}}$ and $B_{k,t}^{\text{recv}})$. By summing them, the total data volume per round is obtained. The total energy consumption for communication at each node is then computed by multiplying the total data volume by the energy consumption per byte (\( E_{\text{byte},k}^{(C)} \)). 

Energy consumption in data communication varies significantly depending on the communication medium, with wired, optical, and wireless methods exhibiting different efficiency. Optical communication is the most energy-efficient, followed by electrical wired networks, while wireless communication (e.g., WiFi, mobile networks) is the most energy-intensive. 

The energy required to transmit one byte of data is summarized in \tablename~\ref{tab:energy_per_byte}. These values highlight the significant disparity in energy efficiency across communication technologies. The information provided by the frontend for \( E_{\text{byte},k}^{(C)} \) is derived from the data presented in \tablename~\ref{tab:energy_per_byte}.

\begin{table}[h]
    \centering
    \caption{Energy Consumption per Byte for Different Communication} Mediums
    \label{tab:energy_per_byte}
    \begin{tabular}{l c}
        \toprule
        \textbf{ Communication} Medium & \textbf{Energy (J/byte)} \\
        \midrule
        Wired (Electrical Signal)~\cite{hu2009semiconductor} & \( 8 \times 10^{-11} \) \\
        Optical Communication ~\cite{tucker2011green} & \( 3.52 \times 10^{-14} \)  \\
        Mobile Network (4G/5G)~\cite{traficom2021} & \( 3.33 \times 10^{-8} \)  \\
        WiFi Communication~\cite{wifi} & \( 5.51 \times 10^{-4} \)  \\
        \bottomrule
    \end{tabular}
\end{table}

The carbon emissions resulting from communication are calculated using Equation~\ref{eq:carbon_emissions}. Similarly, by aggregating the results across all participating nodes, the overall communication-related energy consumption and carbon emissions of the federated system can be obtained. 

All energy consumption and carbon emissions data are transmitted via REST API to TensorBoard, which continuously monitors sustainability metrics throughout the DFL process.

\section{Experimental Evaluation}
\label{sec:exp}
This work employs an experimental approach utilizing the \solution{} framework to evaluate the sustainability of DFL systems systematically. The experiments are designed to address the the research questions defined in Section~\ref{sec:researchmethodology}.

Based on these research questions, this study designs various experimental settings to systematically assess the impact of different DFL configurations and sustainability-aware optimizations.

\subsection{Experiments Setup}
This section describes the experimental setup used to evaluate the sustainability of DFL under different configurations. The experiments systematically analyze various factors, including datasets, model architectures, communication mediums, geographical distribution, network topology, and aggregation algorithms. \tablename~\ref{tab:experiment_summary} summarizes the key experimental parameters.

\begin{table}[h]
    \centering
    \caption{Summary of Experimental Setup}
    \label{tab:experiment_summary}
    \begin{tabular}{@{} p{3.5cm}p{4.5cm}@{}}
        \toprule
        \textbf{Category} & \textbf{Experimental Configurations} \\
        \midrule
        \textbf{Datasets} & MNIST, EMNIST, FashionMNIST, CIFAR10 \\
        \textbf{Models} & MLP, ResNet-9, MobileNetV3 \\
        \textbf{Communication Mediums} & Electrical Signal, Optical Fiber, Mobile Network \\
        \textbf{Geographical Distribution} & Switzerland, Spain \\
        \textbf{Number of Nodes} & 5, 10, 15, 20 \\
        \textbf{Hardware Accelerators} & CPU, GPU \\
        \textbf{Data Distribution} & IID, Non-IID (Dirichlet $\alpha=0.1$) \\
        \textbf{Network Topologies} & Fully Connected, ER ($p=0.5$), Ring \\
        \textbf{Aggregation Algorithms} & FedAvg, Krum, \textbf{\textit{GreenDFL-SA}}, \textbf{\textit{GreenDFL-SN}} \\
        \bottomrule
    \end{tabular}
\end{table}

\paragraph{\textbf{Datasets and Models}}
The experiments utilize MNIST, EMNIST, FashionMNIST, and CIFAR-10, representing different task complexities and dataset sizes. These datasets are widely adopted in the FL research~\cite{thakur2025greenfl}. Moreover, these datasets were selected to provide a representative evaluation across different levels of complexity and data distributions. For \textbf{\textit{RQ1}}, standard benchmark datasets allow measuring energy consumption consistently across the full DFL lifecycle (training, communication, and aggregation).For \textbf{\textit{RQ2}}, datasets with different distributions and varying scales are used to examine how topology, geographic location, and environmental factors influence sustainability. For \textbf{\textit{RQ3}}, more complex datasets such as CIFAR-10 amplify the impact of optimization strategies, making it possible to assess the effectiveness of sustainability-aware algorithms.

\begin{itemize}
    \item \textbf{MNIST}~\cite{lecun1998mnist} dataset is a widely used benchmark for handwritten digit classification in FL and CV)research. It consists of 10 classes, where each sample is a 28×28 grayscale image. The dataset contains 60,000 training samples and 10,000 test samples.  For this task, two different neural network architectures are employed: a three-layer MLP with 256-128-10 hidden units (with \(2.35 \times 10^{5}\) trainable parameters), and a ResNet-9 model (with \(1.6 \times 10^{6}\) trainable parameters)~\cite{he2016resnet}.
    
    \item \textbf{EMNIST}~\cite{cohen2017emnist} dataset is an extension of MNIST, incorporating both digits and handwritten English letters. This work used the "bymerge" configuration for the dataset, which consists of 47 classes. Like MNIST, each sample is a 28×28 grayscale image, but the EMNIST dataset is significantly larger, containing 731,668 training samples and 82,587 test samples. The model architectures used for EMNIST are similar to those employed for MNIST, with modifications to the output layer to accommodate the 47-class classification task.  

    \item \textbf{FashionMNIST}~\cite{xiao2017fashionmnist} dataset a 10-class classification task involving grayscale images of fashion items. It serves as a more challenging alternative to MNIST. The dataset structure is similar to MNIST, with each sample being a 28×28 grayscale image. It includes 60,000 training samples and 10,000 test samples. For this task, the same MLP and ResNet-9 architectures are applied.

    \item \textbf{CIFAR10}~\cite{krizhevsky2009cifar10} dataset is a 10-class classification task involving objects such as animals and vehicles. It presents a higher level of complexity compared to MNIST and FashionMNIST, as each sample is a 32×32 RGB image with three color channels. The dataset consists of 50,000 training samples and 10,000 test samples. To handle the increased complexity, two different convolutional neural network (CNN) architectures are used: MobileNetV3 (with \(1.36 \times 10^{5}\) trainable parameters)~\cite{howard2019searchingmobilenetv3} and ResNet-9  (with \(1.6 \times 10^{6}\) trainable parameters). 
    
This dataset and model selection allows the evaluation of model complexity and dataset difficulty on energy consumption and sustainability in DFL systems.     
\end{itemize}

\paragraph{\textbf{Communication Mediums}}
The choice of communication medium influences the energy consumption of DFL communication, thereby affecting the overall sustainability of the DFL system. To evaluate this impact, the experiments compare the energy consumption and carbon emissions of three different communication mediums: Electrical Signal (Wired Ethernet), Optical Fiber, and Mobile Network (4G).

\paragraph{\textbf{Geographical Distribution}}
The carbon intensity of electricity grids varies significantly across different regions, influencing the carbon emissions of DFL systems. To analyze this effect, the experiment compares DFL deployments in two regions with different carbon intensities: 
\begin{itemize} 
\item \textbf{Spain}: Represents a moderate-carbon-intensity region, with an electricity grid carbon intensity of 217.422 grams of CO$_2$ equivalents per kilowatt-hour (gCO$_2$).  
\item \textbf{Switzerland}: Represents a low-carbon-intensity region, with an electricity grid carbon intensity of 41.279 gCO$_2$.  
\end{itemize}

By comparing the carbon emissions of DFL nodes in these two regions, this study aims to quantify the impact of geographical distribution on the sustainability of DFL. The selection of Switzerland and Spain is motivated by their distinct energy mixes—Switzerland’s grid is largely powered by low-carbon hydropower, while Spain still relies more heavily on fossil fuels despite its growing renewable capacity. This contrast provides a representative proof-of-concept to demonstrate how regional carbon intensity affects DFL sustainability.

\paragraph{\textbf{Federation Size}}
Experiments are conducted with 5, 10, 15, and 20 nodes to assess the effect of federation size on energy consumption and the scalability of the proposed aggregation algorithm.

\paragraph{\textbf{Hardware Accelerators}}
Training is performed on CPU-based computing and GPU-accelerated computing to compare energy efficiency. The experiments are conducted on a device equipped with an AMD EPYC 7702 64-core Processor with a TDP of 200W and an NVIDIA T4 GPU with a TDP of 70W. Each node is virtualized using Docker containers to ensure reproducibility and efficient resource allocation.

\paragraph{\textbf{Data Distribution}}
The distribution of training data across nodes significantly affects the performance of DFL models~\cite{feng2025fedeptailoringattentionheterogeneous}. However, its impact on sustainability, particularly in terms of energy consumption and carbon emissions, remains largely unexplored.

To investigate this relationship, the experiment adopts two different data partitioning strategies:

\begin{itemize}
    \item \textbf{IID (Independent and Identically Distributed)}: Each node receives an evenly distributed subset of the dataset, ensuring uniform data representation across all nodes.
    \item \textbf{Non-IID (Dirichlet Sampling, $\alpha = 0.1$)}: Data is sampled using a Dirichlet distribution with $\alpha = 0.1$, leading to highly skewed and heterogeneous data distributions among nodes.
\end{itemize}

By comparing these two data partitioning methods, this study aims to assess how data heterogeneity influences the sustainability of DFL.

\begin{table*}[h]
    \centering
    \caption{Carbon Emissions (gCO$_2$) and Energy Consumption (kWh) Across Different Datasets with 10 Nodes in Fully Connected Topology}
    \label{tab:energy_carbon}
    \renewcommand{\arraystretch}{1}
    \setlength{\tabcolsep}{4pt}
    \begin{tabular}{lcccccccc}
        \toprule
        \textbf{Dataset} & \textbf{Train CE} & \textbf{Train EC} & \textbf{Agg. CE} & \textbf{Agg. EC} & \textbf{Comm. CE} & \textbf{Comm. EC} & \textbf{Total CE} & \textbf{Total EC} \\
        & (gCO$_2$) & (kWh) & (gCO$_2$) & (kWh) & (gCO$_2$) & (kWh) & (gCO$_2$) & (kWh) \\
        \midrule
        CIFAR10       & 4.047  & 0.019  & 0.534  & 0.002  & \(1.02 \times 10^{-5}\) & \(4.69 \times 10^{-8}\)  & 4.581  & 0.021 \\
        EMNIST        & 6.047  & 0.028  & 0.432  & 0.002  & \(1.74 \times 10^{-5}\) & \(8.02 \times 10^{-8}\)  & 6.479  & 0.030 \\
        FashionMNIST  & 1.300  & 0.006  & 0.413  & 0.002  & \(1.72 \times 10^{-5}\) & \(7.89 \times 10^{-8}\)  & 1.714  & 0.008 \\
        MNIST         & 1.256  & 0.006  & 0.416  & 0.002  & \(1.70 \times 10^{-5}\) & \(7.83 \times 10^{-8}\)  & 1.672  & 0.008 \\
        \bottomrule
    \end{tabular}
    
    \vspace{5pt}
    \small
    \textbf{Abbreviations:} CE: Carbon Emissions, EC: Energy Consumption, Agg.: Aggregation, Comm.: Communication.
\end{table*}

\paragraph{\textbf{Network Topology}}
Network topology defines the communication between nodes in a DFL system and influences the aggregation of models. Different levels of connectivity affect the efficiency of model updates and the overall energy consumption of the system. To analyze its impact on sustainability, the following network topologies are considered:

\begin{itemize}
    \item Fully Connected (Dense): Each node is connected to all others, providing the highest level of communication redundancy and synchronization.
    \item Erdős-Rényi (ER) Random Graph (\( p = 0.5 \)): A probabilistic model where each link exists with probability \( p = 0.5 \), representing a moderately dense network.
    \item Ring Topology (Sparse): Nodes are arranged in a circular manner, with each node connected only to its immediate neighbors.
\end{itemize}

These topologies range from dense to sparse and are used to examine their effects on energy consumption, communication overhead, and overall sustainability in a DFL system.

\paragraph{\textbf{Aggregation}}
Aggregation algorithms influence the energy consumption of the model aggregation process in DFL. To evaluate this impact, three different aggregation strategies are studied:

\begin{itemize}
    \item FedAvg~\cite{mcmahan2017fedavg}: A widely used federated averaging algorithm that computes the weighted average of local models.
    \item Krum~\cite{blanchard2017krum}: A Byzantine-robust aggregation algorithm that selects a single model update closest to the majority of other updates.
    \item \textit{GreenDFL-SA}: An algorithm designed to optimize sustainability by considering energy efficiency during aggregation.
    \item \textit{GreenDFL-SN}: An algorithm optimizes sustainability by considering energy efficiency during the local training phase.
\end{itemize}

All experiments are conducted using 20 aggregation rounds, with each round consisting of 3 local epochs.

\subsection{Analysis of Environmental Impact Across DFL Lifecycle Phases}
\label{sec:rq1}
While the training phase is often seen as the most computation-intensive, it is not guaranteed to be the most energy-consuming. The first experiment investigates which phase in the DFL learning lifecycle contributes the most to environmental sustainability impact.  This experiment is conducted on four datasets using a fully connected topology DFL system with 10 nodes. Communication is performed using Electrical Signal, training is executed on GPU. In this experiment, FedAvg was selected as the baseline aggregation method because it is the most widely adopted and well-studied algorithm in the FL and DFL literature. Besides, FedAvg provides a clear and reproducible baseline. Moreover, MNIST, FashionMNIST, and EMNIST were trained with the MLP model, while the CIFAR-10 dataset was trained with the MobileNetV3 model. All federation nodes are located in Spain, where data distribution follows an IID pattern across nodes.

\tablename~\ref{tab:energy_carbon} presents the total energy consumption and carbon emissions across all nodes for each of the three phases in the DFL lifecycle. The results indicate that the DFL training on the EMNIST dataset consumed the most energy. The differences in energy consumption and carbon emissions across datasets primarily stem from variations in DFL training time. The total learning duration for each dataset is as follows: CIFAR10 takes 14 minutes 46.598 seconds, EMNIST requires 21 minutes 41.565 seconds, FashionMNIST completes in 6 minutes 5.294 seconds, and MNIST finishes in 5 minutes 42.212 seconds. Among these, EMNIST has the longest training time due to its larger dataset size. Although CIFAR10 is the most complex dataset, its smaller data volume results in the second-longest training time. FashionMNIST and MNIST have similar dataset sizes and complexity, leading to nearly identical training durations.

Compared to communication and aggregation, the training phase exhibits the highest energy consumption and carbon emissions. The average local training time for one round on MNIST and FashionMNIST requires approximately 5 seconds per round, whereas EMNIST takes 20 seconds, and CIFAR10 takes 15 seconds per round. In contrast, the aggregation phase takes less than one second for all datasets. Since the computational overhead of training significantly exceeds that of aggregation, the training phase dominates overall energy consumption and carbon emissions.

Although a substantial portion of DFL runtime is spent in the model communication phase, this stage involves minimal computational overhead, resulting in negligible energy consumption compared to training. Consequently, the training phase remains the most energy-intensive stage, contributing the most to overall carbon emissions.

Overall, these findings provide a clear answer to \textbf{\textit{RQ1}}. Across multiple datasets, the training phase is identified as the primary contributor to carbon emissions. Consequently, optimizing the sustainability of the training process should be prioritized in DFL system design. Additionally, energy consumption and emissions from aggregation and communication should not be overlooked. 


\subsection{Factors Influencing the Sustainability of DFL Systems}
\label{sec:rq2}
This section analyzes multiple factors that influence the sustainability of DFL systems, including model architecture, communication medium, geographical distribution of nodes, hardware accelerators, data distribution across nodes, network topology, and federation size. A controlled-variable methodology is adopted, where one factor is varied at a time while others are held constant, to assess its impact on sustainability metrics. The results demonstrate how each factor contributes differently to energy consumption and carbon emissions, providing guidance for sustainable DFL deployment strategies.

\subsubsection{Communication Medium}
This experiment evaluates the impact of different communication media on DFL's energy consumption and carbon emissions. The experiment compares three communication mediums:  Electrical Signal, Optical Fiber, and  Mobile Network. This experiment was conducted on four datasets using a 10-node DFL system with a fully connected topology. 
Training was performed on GPUs, with FedAvg as the aggregation method. 
For the models, MNIST, FashionMNIST, and EMNIST were trained with an MLP, while CIFAR-10 was trained with MobileNetV3. All federation nodes were deployed in Spain with data distributed in an IID manner.
\begin{figure}[h!]
    \centering
    \includegraphics[width=1\linewidth]{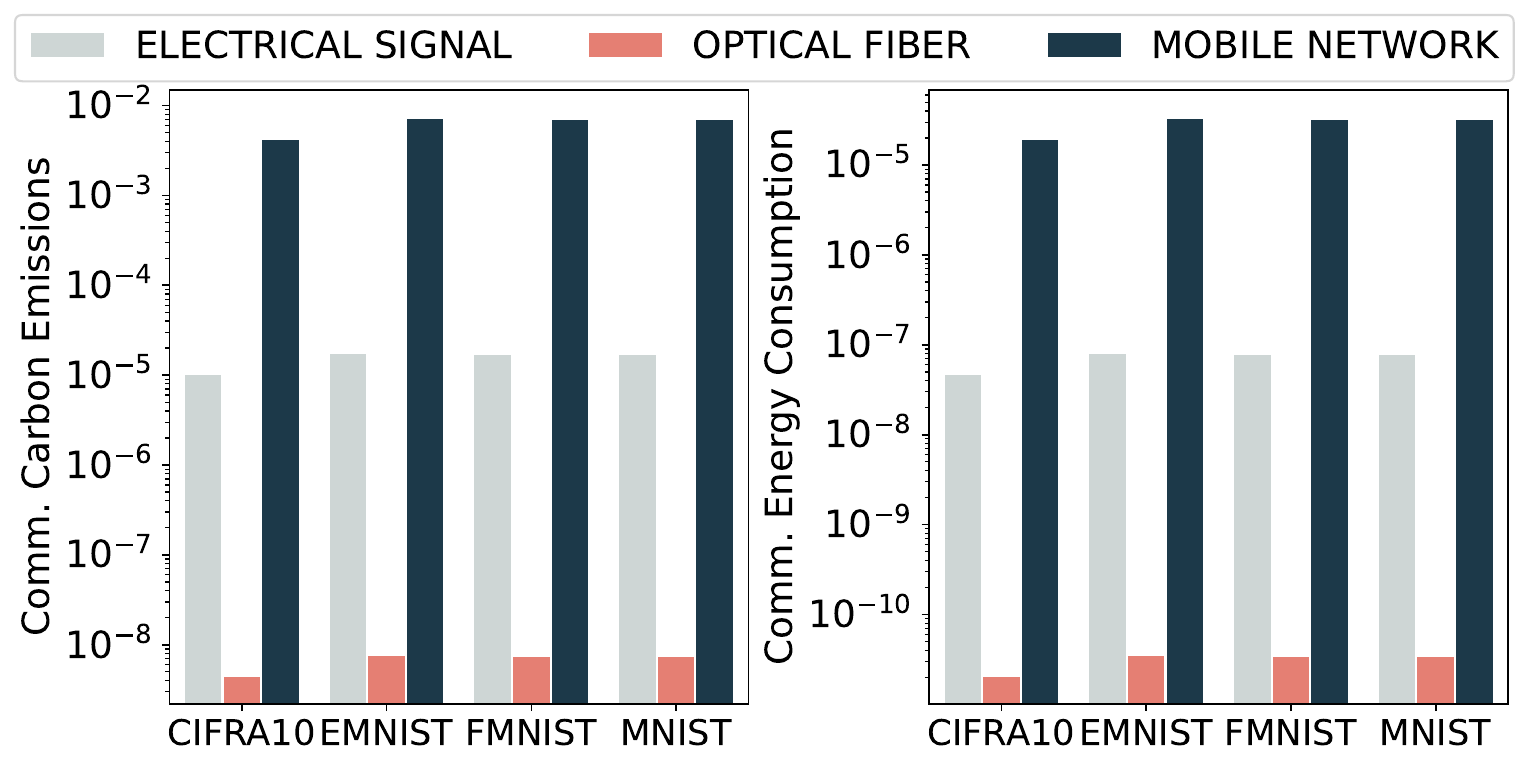}
    \caption{Carbon Emissions (gCO$_2$) and Energy Consumption (kWh) During the Communication Phase With Various Communication Medium Across Different Datasets (Log-scaled)}
    \label{fig:energy_carbon_comm}
\end{figure}

The communication medium mainly affects energy consumption during the communication phase of the DFL system. \figurename~\ref{fig:energy_carbon_comm} illustrates the energy consumption and carbon emissions during the communication phase across four datasets using three different communication mediums.

The results indicate that optical fiber, which has the lowest per-byte energy consumption, results in the lowest communication energy consumption and carbon emissions under the same setup. In contrast, mobile communication exhibits the highest per-byte energy consumption, leading to the highest communication energy consumption and emissions. However, even when using mobile communication, the total carbon emissions from the communication phase over 20 rounds remain relatively low, contributing only approximately 0.01 gCO$_2$e across all four datasets.

\subsubsection{Geographical Distribution}
\begin{figure}[h!]
    \centering
    \includegraphics[width=1\linewidth]{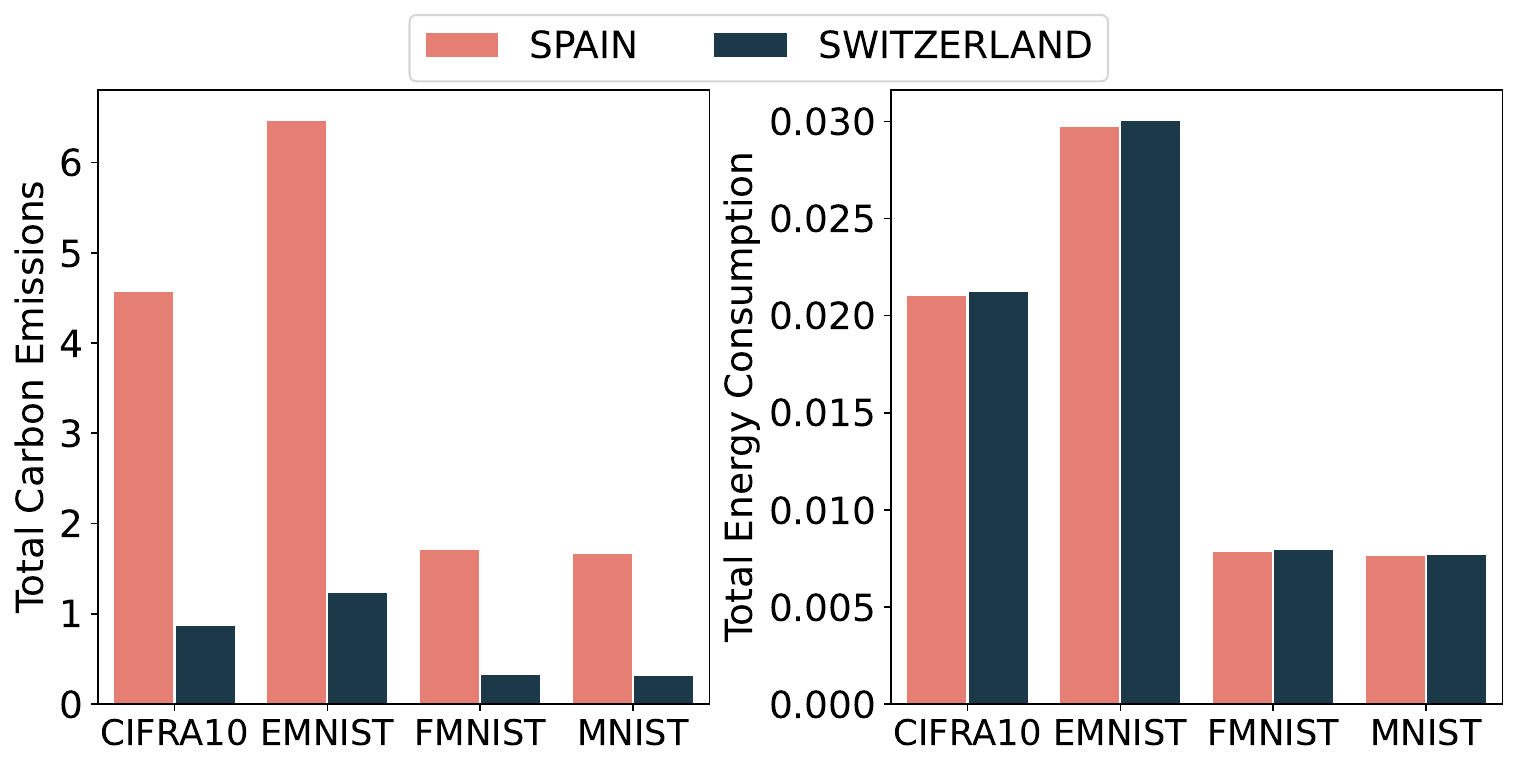}
    \caption{Total Carbon Emissions (gCO$_2$) and Energy Consumption (kWh) of DFL Systems in Spain and Switzerland}
    \label{fig:energy_carbon_location}
\end{figure}

The geographic distribution of nodes affects the carbon intensity of the electricity grid they utilize, thereby influencing the overall carbon emissions of the DFL system. This experiment compares the energy consumption and carbon emissions of DFL systems deployed in Spain and Switzerland, as illustrated in \figurename~\ref{fig:energy_carbon_location}. This experiment was conducted on four datasets using a 10-node DFL system with a fully connected topology implemented on the Electrical Signal testbed. Training was executed on GPUs with FedAvg as the aggregation method. MNIST, FashionMNIST, and EMNIST were trained with an MLP, while CIFAR-10 was trained with MobileNetV3. All federation nodes were deployed with data distributed in an IID manner.

The comparison between Switzerland and Spain highlights the effect of regional energy mixes on DFL sustainability. Switzerland relies largely on renewable sources, particularly hydropower, leading to low carbon intensity, whereas Spain, despite increasing use of wind and solar, still maintains a higher share of fossil fuels, resulting in higher carbon intensity.

Under the same configuration, the geographic distribution of nodes does not impact energy consumption; however, it significantly affects carbon emissions. Although the DFL systems deployed in Switzerland and Spain consume a similar amount of energy under comparable settings, the difference in grid carbon intensity results in substantial disparities in carbon emissions. Specifically, Switzerland’s electricity grid has a carbon intensity of 41.279~gCO$_2$/kWh, which is about one-fourth of Spain’s 217.422~gCO$_2$/kWh. According to Equation~\ref{eq:carbon_emissions}, the total carbon emissions are therefore approximately one-fourth in Switzerland compared to Spain. These results suggest that optimizing the geographic distribution of nodes can effectively reduce the environmental impact of DFL systems. In particular, selecting nodes in regions with lower carbon intensity for training further minimizes overall carbon emissions.

\subsubsection{Hardware Accelerator}
\begin{figure}[h!]
    \centering
    \includegraphics[width=1\linewidth]{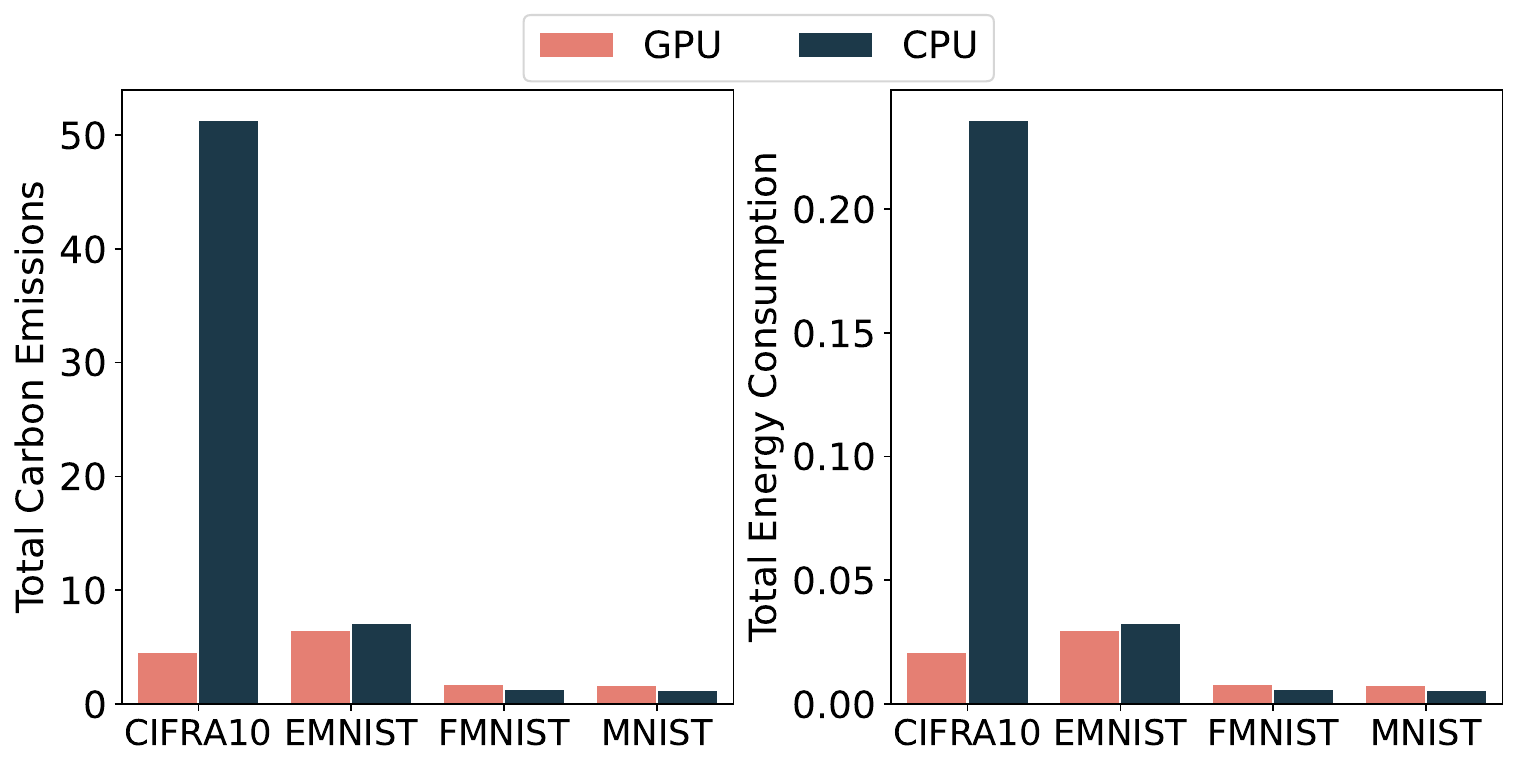}
    \caption{Total Carbon Emissions (gCO$_2$) and Energy Consumption (kWh) of DFL Systems with Using GPU and CPU Accelerators}
    \label{fig:hardware_energy}
\end{figure}

The choice of hardware accelerator significantly affects the sustainability of a DFL system by influencing computational efficiency during local training and aggregation. Due to their superior performance in tensor computations, GPUs can significantly reduce training time compared to CPUs. Additionally, GPUs offer better power management, resulting in lower overall energy consumption. This experiment was conducted on a 10-node DFL system with a fully connected topology by using electrical signal. FedAvg was employed as the aggregation method. MNIST, FashionMNIST, and EMNIST were trained with an MLP, while CIFAR-10 was trained with MobileNetV3. All federation nodes were deployed in Spain. \figurename~\ref{fig:hardware_energy} compares the energy consumption and carbon emissions of DFL systems using CPUs and GPUs as accelerators.

For the EMNIST, FashionMNIST, and MNIST datasets, training with the MLP model exhibits similar learning durations on both CPU and GPU, resulting in similar energy consumption. However, for the CIFAR10 dataset, training on the CPU takes nearly 80 minutes while operating at full capacity. Consequently, its energy consumption is approximately ten times that of GPU-based training, leading to a nearly tenfold increase in equivalent CO$_2$ emissions.

The results show that DFL systems utilizing GPU accelerators achieve better energy efficiency and lower carbon emissions under the same configuration. In contrast, CPU-only systems require longer training times, leading to higher energy consumption and carbon emissions.

\subsubsection{Model Architecture}
More complex models often yield better performance but require higher computational resources. This experiment was conducted on a 10-node DFL system with a fully connected topology located in Spain, using GPUs for training and FedAvg as the aggregation algorithm. This experiment evaluates the effect of simple models (MLP for MNIST, FashionMNIST, and EMNIST; MobileNet for CIFAR10) versus a more complex model (ResNet-9) for all four datasets on the sustainability of DFL systems. 

\begin{figure}[h!]
    \centering
    \includegraphics[width=1\linewidth]{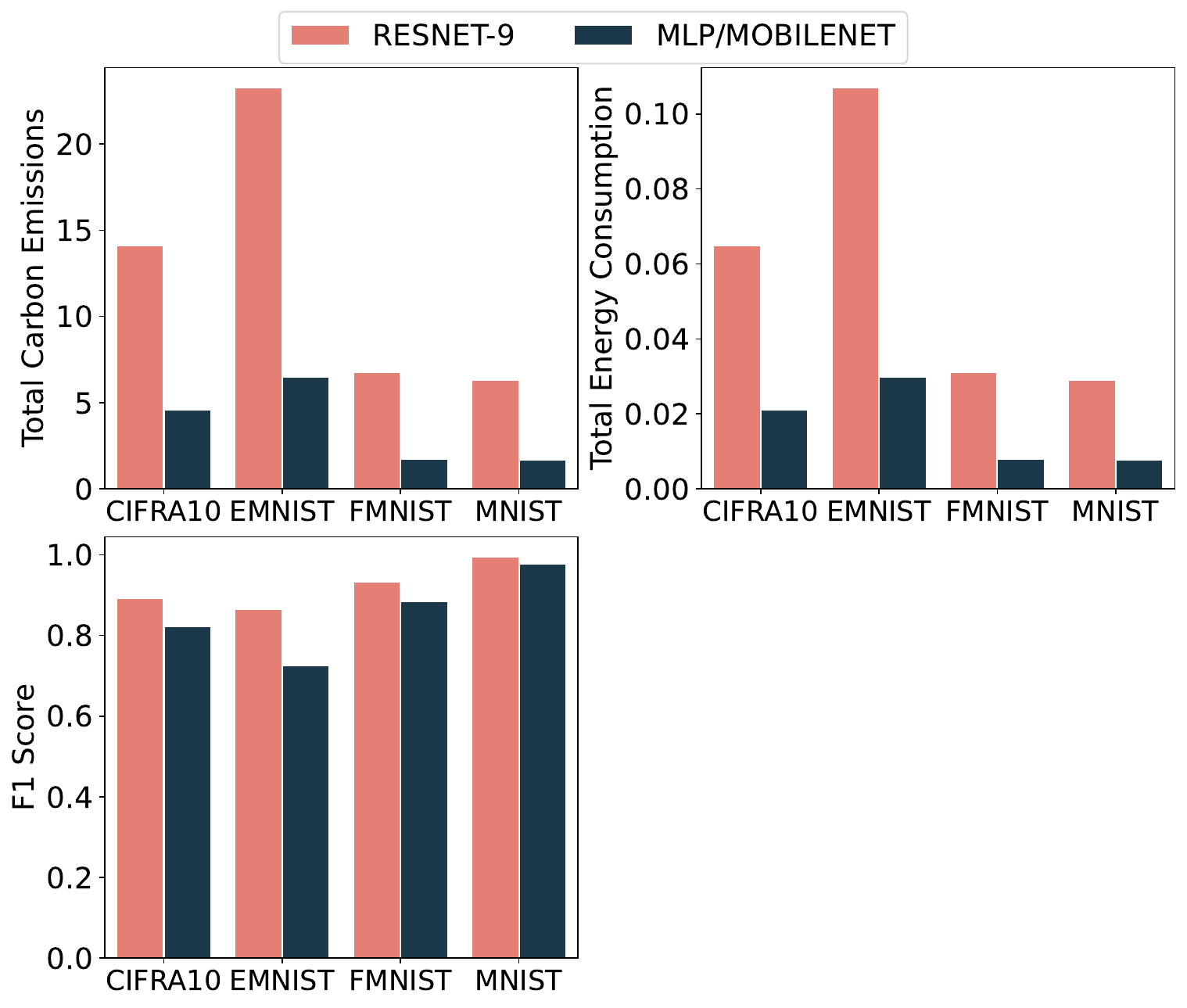}
    \caption{Total Carbon Emissions (gCO$_2$), Energy Consumption (kWh) and Test F1 Score of DFL Systems with Different Model Architectures}
    \label{fig:model_performance}
\end{figure}

\figurename~\ref{fig:model_performance} compares different model architectures' carbon emissions, energy consumption, and F1 score. The results indicate that while ResNet-9 improves the F1 score in all datasets, it also leads to higher carbon emissions. This is due to its significantly larger parameter count (\(1.6 \times 10^6\)), approximately seven times that of MLP and MobileNet models. Consequently, ResNet-9 has a higher computational density and requires longer training time, resulting in increased energy consumption and carbon emissions.

\subsubsection{Network Topology}
\begin{figure}[h!]
    \centering
    \includegraphics[width=1\linewidth]{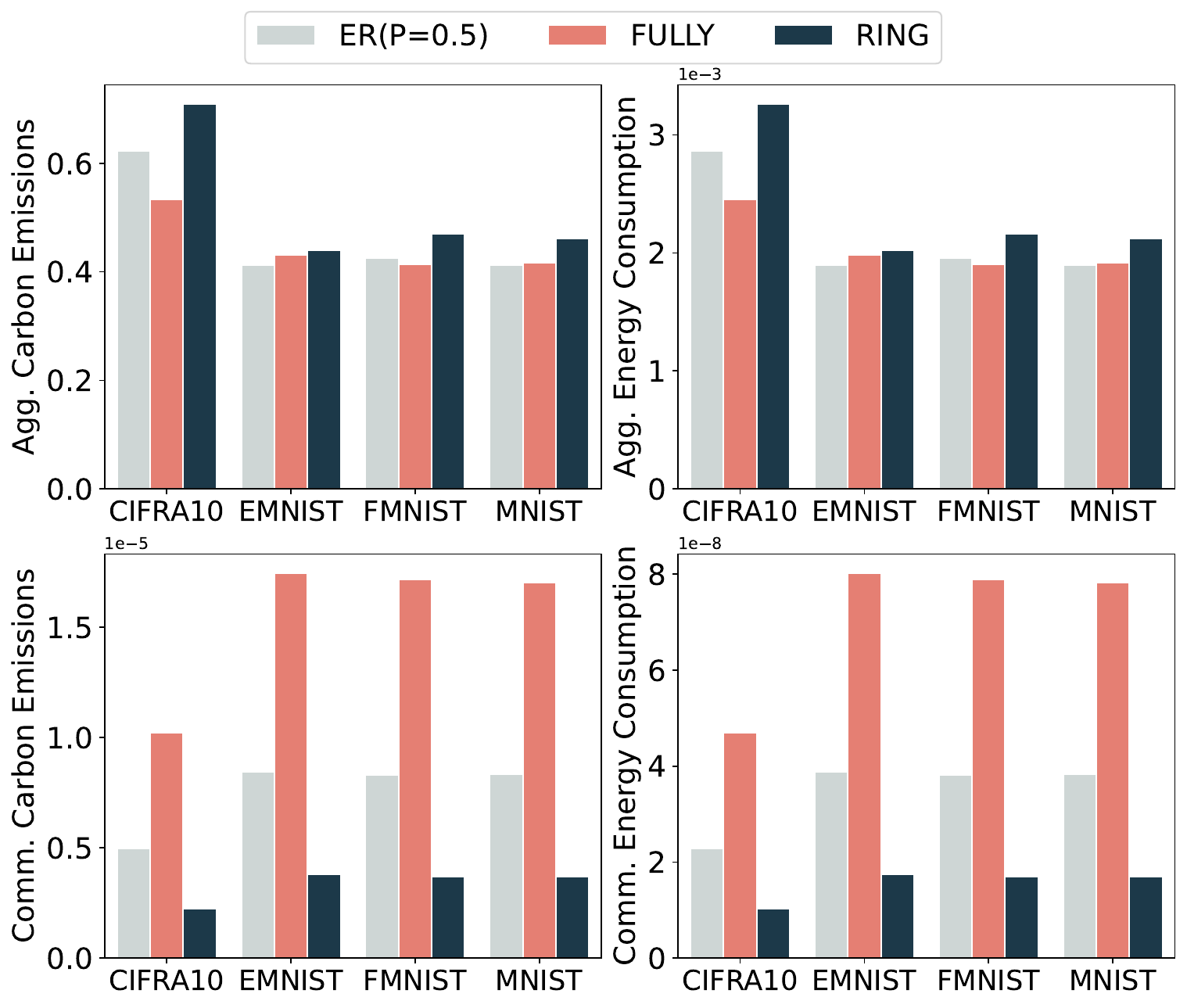}
    \caption{Total Carbon Emissions (gCO$_2$) and Energy Consumption (kWh) of DFL Systems with Different Network Topologies}
    \label{fig:topology_impact}
\end{figure}

Network topology determines how models are transmitted between nodes and how many models are aggregated, influencing energy consumption and carbon emissions during the communication and aggregation phases. To assess this impact, the experiment evaluated three topologies: fully connected, ER (\( p = 0.5 \)), and ring topology. This experiment was conducted on four datasets using a 10-node DFL system. 
FedAvg was employed as the aggregation method. MNIST, FashionMNIST, and EMNIST were trained with an MLP, while CIFAR-10 was trained with MobileNetV3. All federation nodes were deployed in Spain with data distributed in an IID manner. \figurename~\ref{fig:topology_impact} shows the energy consumption and carbon emissions in the communication and aggregation phases across four datasets.

The results indicate that under different network topologies, the energy consumption and carbon emissions during the aggregation phase remain similar. This suggests that although sparse topologies, such as the ring topology, require fewer models to be processed during aggregation, the computational time for aggregation is relatively short, and the overall computational overhead remains low. Consequently, energy consumption during aggregation does not show significant differences between sparse and dense networks.

However, the communication phase exhibits differences. The total number of model exchange in a DFL system per round is theoretically twice the number of edges in the network. In a fully connected network with \( N \) nodes, the total number of edges is \( {N(N-1)}/{2} \), meaning that in each round, the system transmits \( N(N-1) \) model updates. In an ER random graph with \( p = 0.5 \), the expected number of edges is \( {N(N-1)}/{4} \), resulting in \( {N(N-1)}/{2} \) model exchange per round. In contrast, a ring topology has exactly \( N \) edges, leading to only \( 2N \) model exchange per round.

In the 10-node DFL system used in this experiment, a fully connected topology required 90 model exchange per round, while the ER topology required 45, and the ring topology only 20. Experimental results confirm this theoretical analysis. Using CIFAR10 as an example, over 20 training rounds, the fully connected topology consumed \( 4.69439 \times 10^{-8} \) kWh in communication, while the ER topology consumed \( 2.29197 \times 10^{-8} \) kWh, approximately half of the fully connected network’s consumption. The ring topology exhibited the lowest energy consumption, around \( 1.02738 \times 10^{-8} \) kWh, aligning with theoretical expectations.

\begin{figure}[h!]
    \centering
    \includegraphics[width=1\linewidth]{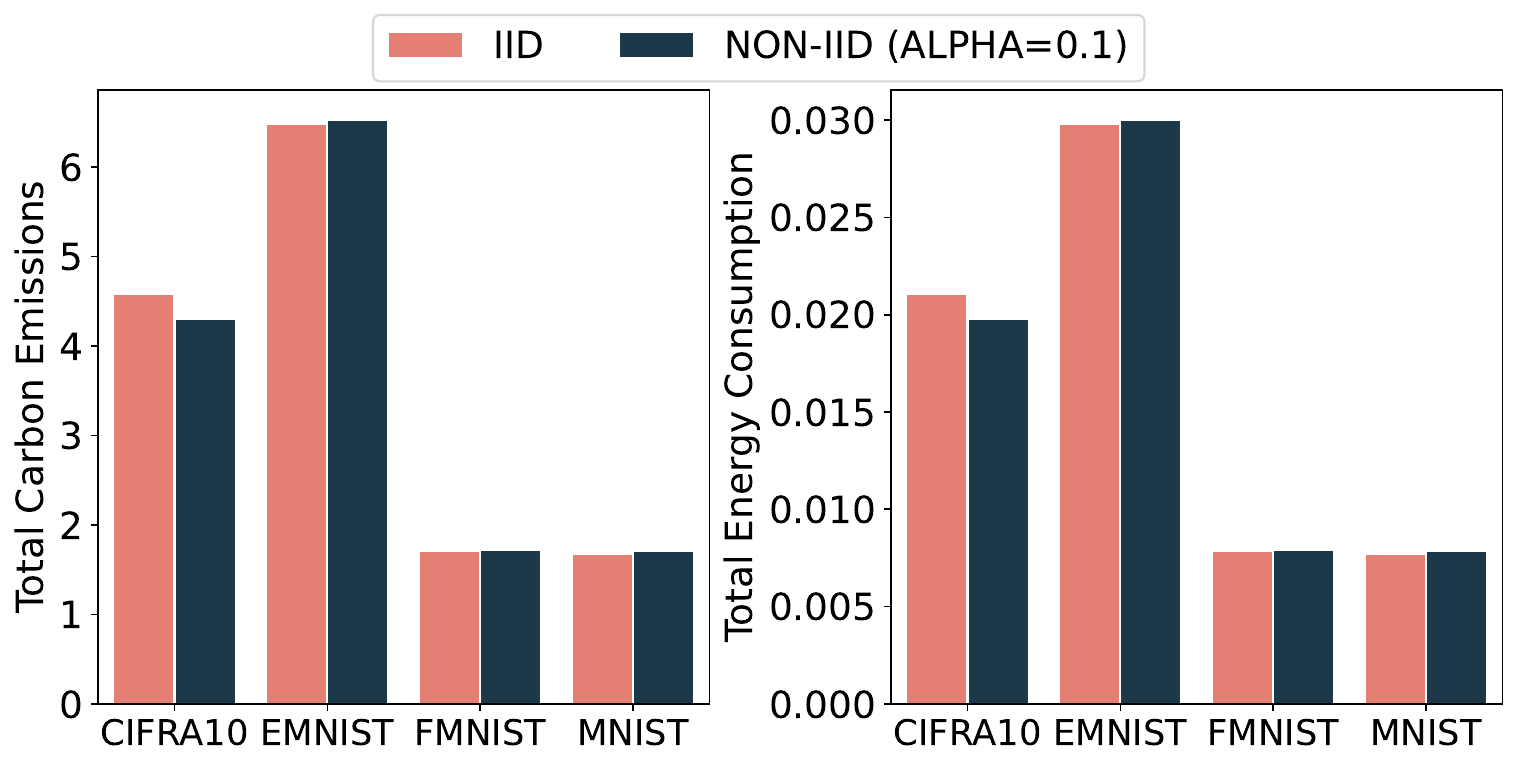}
    \caption{Total Carbon Emissions (gCO$_2$), Energy Consumption (kWh) and Test F1 Score of DFL Systems with Different Data Distribution}
    \label{fig:data_distribution}
\end{figure}

\subsubsection{Data Distribution}
\figurename~\ref{fig:data_distribution} presents the carbon emissions and energy consumption under IID and non-IID \( \alpha = 0.1 \) distributions. This experiment was conducted on four datasets using a 10-node DFL system with fully connected topology. FedAvg was employed as the aggregation method. MNIST, FashionMNIST, and EMNIST were trained with an MLP, while CIFAR-10 was trained with MobileNetV3. All federation nodes were deployed in Spain. Under non-IID conditions, the overall energy consumption and carbon emissions of the system remain similar to those observed under IID settings when using the same model and aggregation algorithm. This suggests that data distribution has a minimal impact on the environmental sustainability of DFL.

\subsubsection{Federation Size}
\begin{figure}[h!]
    \centering
    \includegraphics[width=1\linewidth]{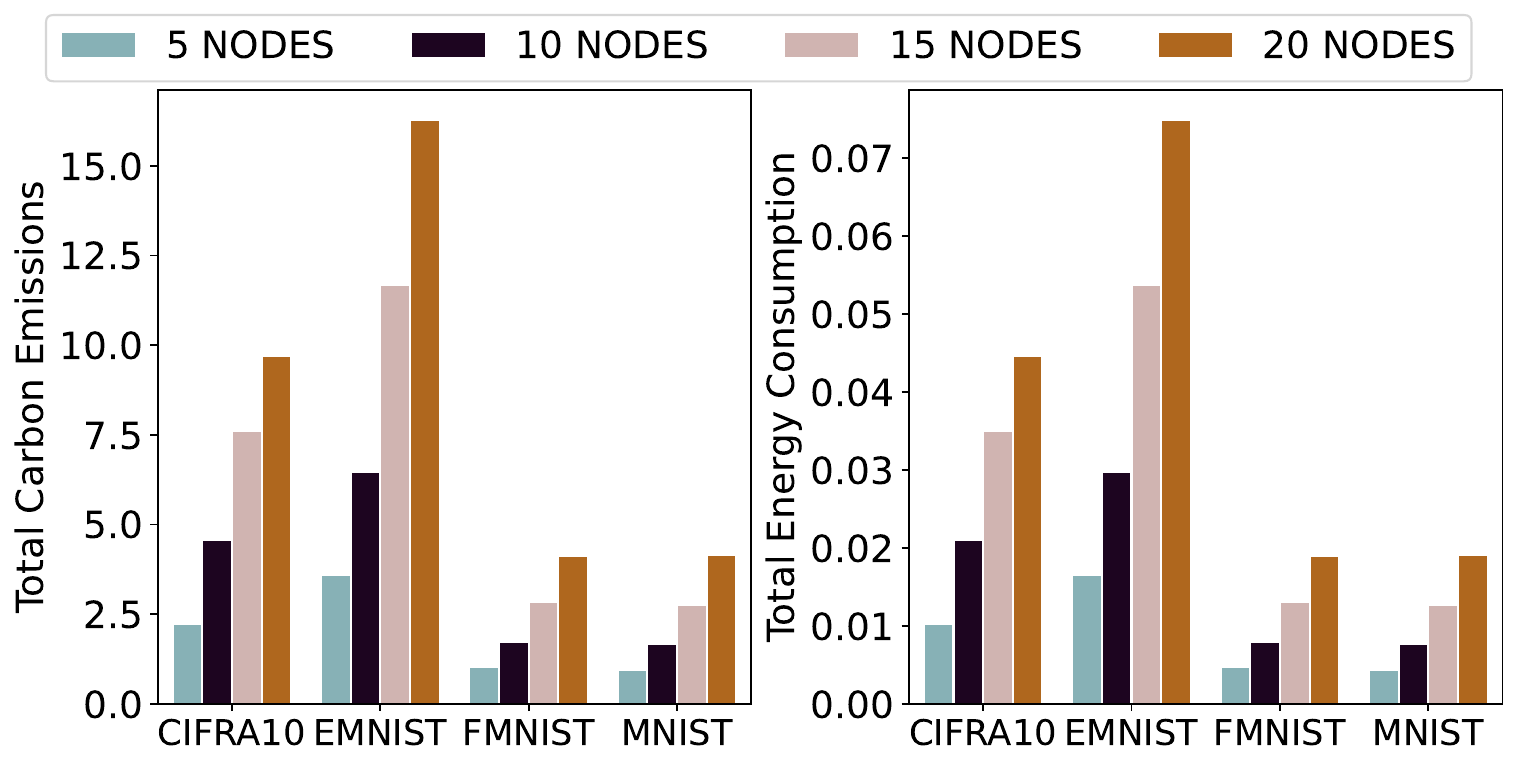}
    \caption{Total Carbon Emissions (gCO$_2$) and Energy Consumption (kWh) of DFL Systems with Different Federation Sizes}
    \label{fig:federation_size}
\end{figure}

This experiment evaluates the differences in energy consumption and carbon emissions when the number of participating nodes is varied between 5, 10, 15, and 20. This experiment was conducted on fully connected topology. FedAvg was employed as the aggregation method. MNIST, FashionMNIST, and EMNIST were trained with an MLP, while CIFAR-10 was trained with MobileNetV3. All federation nodes were deployed in Spain. The results are presented in \figurename~\ref{fig:federation_size}.

As the number of nodes in the federation increases, the energy consumption and carbon emissions per node remain similar. However, the total energy consumption of the system increases proportionally with the number of participating nodes. A larger number of nodes in training leads to higher overall energy consumption and carbon emissions.

In conclusion, factors such as geographical distribution, hardware accelerators, model architecture, and federation size significantly impact DFL sustainability. While communication medium and network topology influence carbon emissions during the communication phase, their overall impact on the sustainability of the DFL system is relatively small. However, data distribution has minimal effect on the system’s sustainability. These insights provide an answer to \textit{\textbf{RQ2}}.

\begin{table*}[h]
    \centering
    \caption{Comparison of Sustainability Metrics and Model Performance Across Aggregation Algorithms}
    \label{tab:sustainability_comparison}
    \resizebox{\textwidth}{!}{%
    \renewcommand{\arraystretch}{1}
    \setlength{\tabcolsep}{4pt}
    \begin{tabular}{l l c c c c c c c c c}
        \toprule
        \textbf{Dataset} & \textbf{Alg.} & \textbf{Train. CE} & \textbf{Train. EC} & \textbf{Agg. CE} & \textbf{Agg. EC} & \textbf{Comm. CE} & \textbf{Comm. EC} & \textbf{Total CE} & \textbf{Total EC} & \textbf{F1 Score} \\
        & & (gCO$_2$) & (kWh) & (gCO$_2$) & (kWh) & (gCO$_2$) & (kWh) & (gCO$_2$) & (kWh) & \\
        \midrule
        \multirow{4}{*}{CIFAR10} 
        & FedAvg  & 4.047  & 0.019  & 0.534  & 0.002  & 1.021 $\times 10^{-5}$  & 4.694 $\times 10^{-8}$  & 4.581  & 0.021  & \textbf{0.822} \\
        & Krum    & 4.016  & 0.018  & 1.283  & 0.006  & 1.019 $\times 10^{-5}$  & 4.688 $\times 10^{-8}$  & 5.299  & 0.024  & 0.785 \\
        & \textbf{\textit{GreenDFL-SA}}      & 4.068  & 0.019  & \textbf{0.497}  & \textbf{0.002}  & 1.022 $\times 10^{-5}$  & 4.702 $\times 10^{-8}$  & 4.565  & 0.021  & 0.817 \\
        & \textbf{\textit{GreenDFL-SN}}      & \textbf{2.761}  & \textbf{0.013}  & 0.674  & 0.003  & \textbf{1.017 $\times 10^{-5}$}  & \textbf{4.675 $\times 10^{-8}$}  & \textbf{3.435}  & \textbf{0.016}  & 0.785 \\
        \midrule
        \multirow{4}{*}{EMNIST} 
        & FedAvg  & 6.047  & 0.028  & 0.432  & 0.002  & 1.743 $\times 10^{-5}$  & 8.017 $\times 10^{-8}$  & 6.479  & 0.030  & \textbf{0.726} \\
        & Krum    & 6.069  & 0.028  & 0.807  & 0.004  & 1.740 $\times 10^{-5}$  & 8.003 $\times 10^{-8}$  & 6.876  & 0.032  & 0.622 \\
        & \textbf{\textit{GreenDFL-SA}}      & 6.090  & 0.028  & 0.410  & 0.002  & 1.738 $\times 10^{-5}$  & 7.994 $\times 10^{-8}$  & 6.500  & 0.030  & 0.718 \\
        & \textbf{\textit{GreenDFL-SN}}      & \textbf{5.612}  & \textbf{0.026}  & 0.448  & 0.002  & \textbf{1.737 $\times 10^{-5}$}  & \textbf{7.991 $\times 10^{-8}$}  & \textbf{6.060}  & \textbf{0.028}  & 0.713 \\
        \midrule
        \multirow{4}{*}{FashionMNIST} 
        & FedAvg  & 1.300  & 0.006  & 0.413  & 0.002  & 1.715 $\times 10^{-5}$  & 7.890 $\times 10^{-8}$  & 1.714  & 0.008  & \textbf{0.884} \\
        & Krum    & 1.324  & 0.006  & 0.895  & 0.004  & 1.698 $\times 10^{-5}$  & 7.808 $\times 10^{-8}$  & 2.219  & 0.010  & 0.851 \\
        & \textbf{\textit{GreenDFL-SA}}      & 1.304  & 0.006  & \textbf{0.397}  & \textbf{0.002}  & \textbf{1.692 $\times 10^{-5}$}  & \textbf{7.781 $\times 10^{-8}$}  & 1.701  & 0.008  & 0.881 \\
        & \textbf{\textit{GreenDFL-SN}}      & \textbf{1.155}  & \textbf{0.005}  & 0.467  & 0.002  & 1.700 $\times 10^{-5}$  & 7.817 $\times 10^{-8}$  & \textbf{1.623}  & \textbf{0.007}  & 0.873 \\
        \midrule
        \multirow{4}{*}{MNIST} 
        & FedAvg  & 1.256  & 0.006  & 0.416  & 0.002  & 1.702 $\times 10^{-5}$  & 7.830 $\times 10^{-8}$  & 1.672  & 0.008  & \textbf{0.978} \\
        & Krum    & 1.227  & 0.006  & 0.820  & 0.004  & 1.698 $\times 10^{-5}$  & 7.811 $\times 10^{-8}$  & 2.047  & 0.009  & 0.961 \\
        & \textbf{\textit{GreenDFL-SA}}      & 1.269  & 0.006  & \textbf{0.392}  & \textbf{0.002}  & 1.698 $\times 10^{-5}$  & 7.810 $\times 10^{-8}$  & 1.661  & 0.008  & 0.976 \\
        & \textbf{\textit{GreenDFL-SN}}      & \textbf{1.056}  & \textbf{0.005}  & 0.422  & 0.002  & \textbf{1.697 $\times 10^{-5}$}  & \textbf{7.806 $\times 10^{-8}$}  & \textbf{1.478}  & \textbf{0.007}  & 0.971 \\
        \bottomrule
    \end{tabular}}
    \small
    \textbf{Abbreviations:} CE: Carbon Emissions, EC: Energy Consumption, Agg.: Aggregation, Comm.: Communication
\end{table*}

\subsection{Analysis of \textit{GreenDFL-SA} and \textit{GreenDFL-SN}}
\label{sec:rq3}
This experiment evaluates the proposed \textit{GreenDFL-SA} and \textit{GreenDFL-SN} algorithms in comparison with FedAvg and Krum. This experiment was conducted on a 10-node DFL system with IID data partitioning and a fully connected network topology. MNIST, FashionMNIST, and EMNIST were trained with an MLP, while CIFAR-10 was trained with MobileNetV3. All federation nodes were deployed in Spain. The evaluation focuses on three key aspects: energy consumption, carbon emissions, and model performance (Test F1 Score).

\tablename~\ref{tab:sustainability_comparison} compares the sustainability metrics and model performance across four training algorithms. The results indicate that the test F1 scores of all four algorithms are similar, demonstrating that the proposed \textit{GreenDFL-SA} and \textit{GreenDFL-SN} methods do not compromise model performance.

Regarding sustainability metrics, FedAvg, Krum, and the \textit{GreenDFL-SA} exhibit comparable energy consumption and carbon emissions. This similarity arises because these three methods differ mainly in the aggregation phase, which contributes relatively little to the system’s overall energy consumption. Consequently, variations in aggregation complexity do not significantly impact total energy usage, particularly for lightweight models such as MLP on MNIST, FashionMNIST, and EMNIST datasets. However, at the aggregation phase level, Krum exhibits higher energy consumption due to its increased computational complexity compared to FedAvg. The \textit{GreenDFL-SA} reduces energy consumption during aggregation, contributing to improved sustainability. The \textit{GreenDFL-SN} algorithm, by dynamically excluding energy-intensive nodes, achieves the best overall energy efficiency, with its benefits becoming more pronounced for complex tasks such as MobileNet on CIFAR-10.

In summary, the proposed \textit{GreenDFL-SA} and \textit{GreenDFL-SN} algorithms reduce energy consumption and carbon emissions at different stages of the DFL lifecycle, thereby enhancing the environmental sustainability of DFL systems. These findings provide a positive answer to \textit{\textbf{RQ3}}, demonstrating the potential for sustainability-aware optimization in decentralized learning.

\section{Discussion}
\label{sec:discussion}
This work constructs an environmental sustainability assessment framework for DFL systems and employs empirical research to identify key influencing factors. The experiments also demonstrate that optimizing aggregation and node selection strategies can enhance the environmental sustainability of DFL. This section presents best practices for improving the sustainability of DFL systems.

\subsection{Balancing Model Performance and Sustainability}
Achieving a balance between model performance and sustainability is challenging. Optimizing model architecture by selecting a model with a sufficient yet minimal number of parameters appears to be a reasonable approach. However, determining the optimal model size before deployment is complex, and such optimization often becomes a post-training consideration.

The proposed \textit{GreenDFL-SN} algorithm offers an in-training strategy to effectively reduce DFL carbon emissions. However, its effectiveness relies on all nodes' accurate and timely reporting of sustainability metrics. In decentralized environments, this information may not always be reliable, limiting the applicability of the approach to scenarios where all nodes are honest. Another straightforward strategy to reduce DFL energy consumption and carbon emissions is adopting the early stopping mechanism.

\begin{figure}[h!]
    \centering
    \includegraphics[width=1\linewidth]{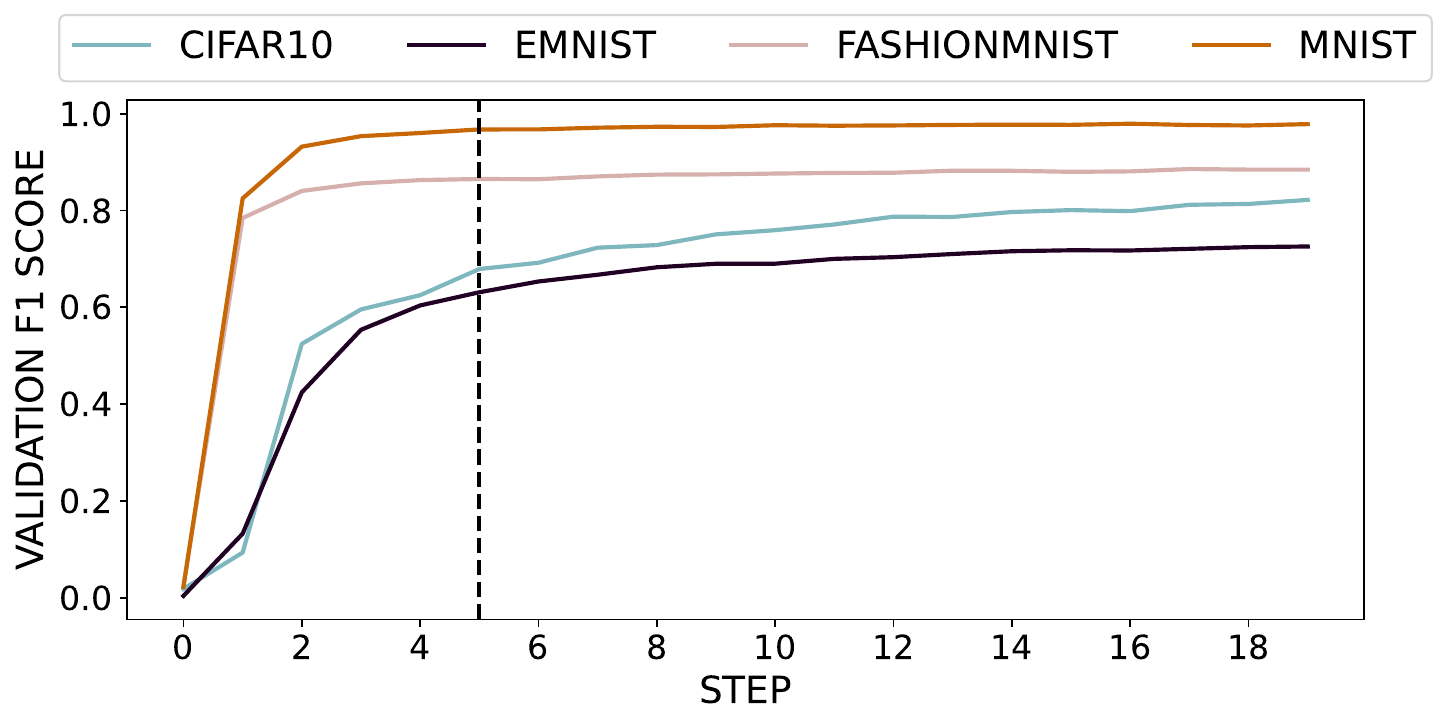}
    \caption{Validation F1 Score Across Different Datasets}
    \label{fig:early_stopping}
\end{figure}

Typically, DFL systems follow a predetermined number of training rounds, regardless of whether the model has already converged. The early stopping strategy allows nodes to monitor their validation metrics, such as loss or accuracy, over multiple rounds and terminate training when the improvement falls within a predefined threshold. \figurename~\ref{fig:early_stopping} illustrates a DFL system's validation loss and sustainability metrics on the MNIST dataset across training rounds. The results indicate that although the model converges by round 5, training continues for the predetermined 20 rounds. If early stopping were applied at round 7, energy consumption could be reduced by approximately 60\%. Therefore, automated convergence detection mechanisms, such as early stopping, can significantly improve DFL systems' energy efficiency and environmental sustainability. 

While early stopping clearly improves energy efficiency, it also introduces trade-offs that deserve further discussion. Early stopping may risk underfitting if validation metrics fluctuate or plateau temporarily, especially in heterogeneous data distributions common in DFL. This risk is amplified for complex datasets and tasks. For example, as shown in the \figurename~\ref{fig:early_stopping} CIFAR-10 results, the model performance continues to improve even after 12 rounds, suggesting that applying early stopping too aggressively may hinder full convergence.

\subsection{Utilization of Renewable Energy}
\begin{figure}[h!]
    \centering
    \includegraphics[width=1\linewidth]{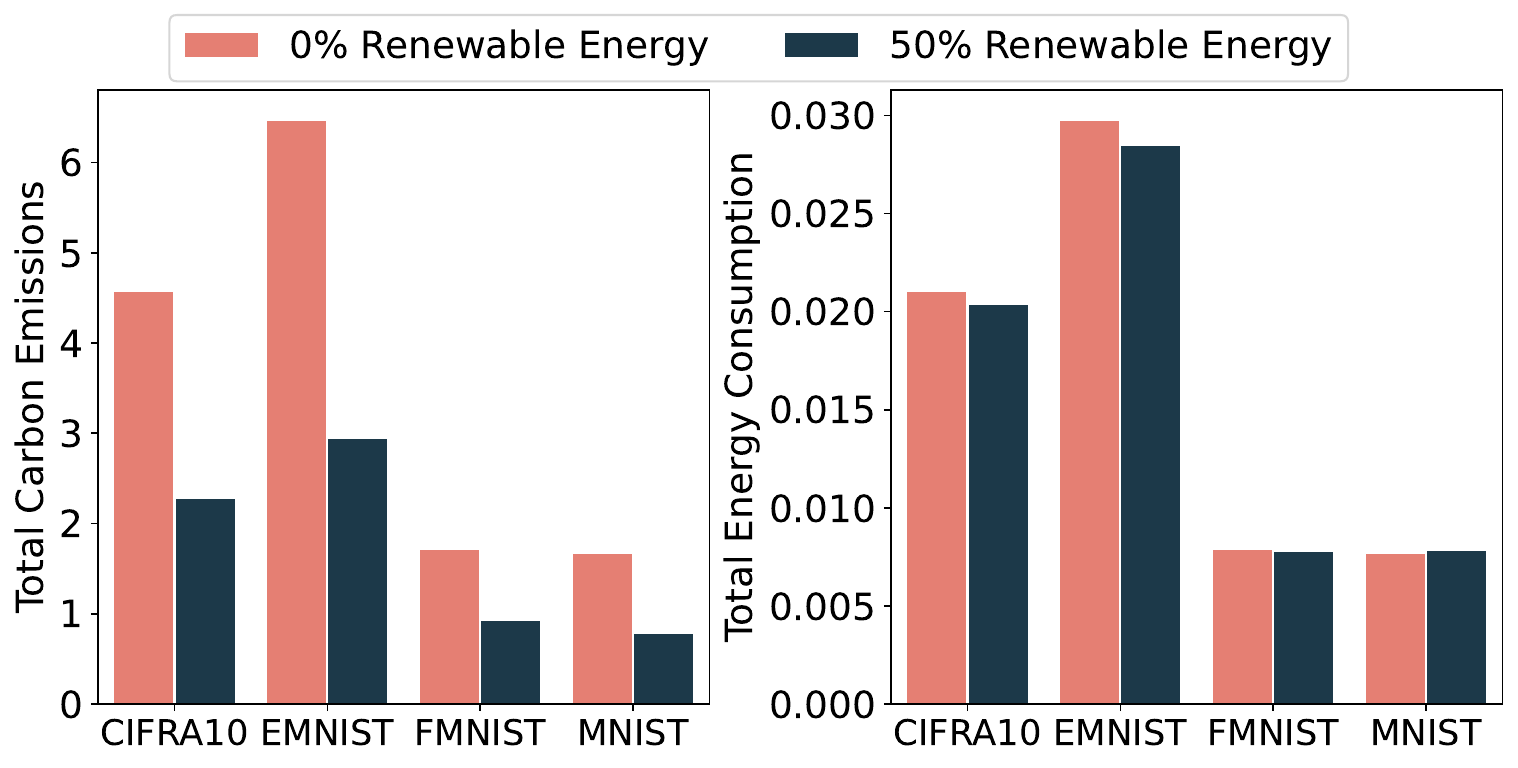}
    \caption{Total Carbon Emissions (gCO$_2$) and Energy Consumption (kWh) of DFL Systems with Different Local Renewable Energy Ratio}
    \label{fig:renewable_energy}
\end{figure}

In the conducted experiments, it was assumed that all nodes exclusively relied on grid electricity. However, many data centers and households are generating their own renewable energy as an alternative power source. As shown in \figurename~\ref{fig:renewable_energy}, when nodes utilized 50\% locally generated renewable energy, the system's total carbon emissions were reduced by approximately 50\%. Investing in and adopting local clean energy sources reduces electricity costs and enhances environmental sustainability.

\section{Threats to Validity}
\label{sec:threatstovalidity}
This section discusses the threats to validity, including internal, external, construct, and conclusion validity.

\subsection{Internal Validity}
The measurement of energy consumption may be influenced by the limitations of GPU power sampling tools, such as sampling frequency or reporting accuracy, while CPU and communication overheads may be underestimated. To mitigate this, frequent runtime GPU power sampling was applied instead of relying on static TDP values, and communication metrics were explicitly incorporated into the overall energy model. 

The estimation of carbon emissions depends on regional carbon intensity values and renewable energy ratios, which represent approximations and may not fully capture the actual local energy mix; this threat was mitigated by using publicly available datasets from trusted energy agencies. 

Although the experiments followed a controlled-variable methodology, unobserved system-level factors (e.g., scheduling behavior or hardware-specific optimizations) could still affect the outcomes.

\subsection{External Validity}
The experiments were conducted on a limited number of datasets (MNIST, FashionMNIST, EMNIST, CIFAR-10) and models (MLP, MobileNetV3, and ResNet-9), which may not capture the full diversity of real-world applications of DFL. Geographical experiments compared only Spain and Switzerland, leaving other regional energy profiles unexplored. These factors constrain the generalizability of the findings. To mitigate this, this work considers tasks of varying complexity, datasets with different sample sizes, and models of different complexity, which strengthens the applicability of the results across diverse DFL scenarios, providing reasonable representativeness.

\subsection{Construct Validity}
The work uses energy consumption and carbon emissions as the primary sustainability metrics. While these indicators capture important aspects of environmental impact, they may not fully reflect broader sustainability dimensions such as hardware lifecycle costs or embodied carbon in devices. In addition, the estimation of GPU power consumption relies on sampled runtime measurements, which may introduce minor deviations compared to true energy usage. Despite these limitations, the chosen metrics align with prior work on Green AI and FL sustainability, providing a consistent basis for evaluation.

\subsection{Conclusion Validity}
The validity of the conclusions depends on the correctness of the experimental design and the consistency of the evaluation metrics. While absolute values of energy consumption and carbon emissions may vary depending on hardware and environmental settings, the relative comparisons across algorithms, datasets, and topologies remain valid. The use of multiple datasets, models, and network configurations helps ensure that the observed trends are not artifacts of a particular setup, thereby supporting the robustness of the conclusions.

\section{Summary and Future Work}
\label{sec:conclusion}
This work presents a comprehensive framework, called \solution{}, for assessing the environmental sustainability of DFL systems. Through empirical analysis, it investigates the impact of key factors such as model architecture, hardware accelerators, communication medium, data distribution, network topology, and federation size on sustainability. The results demonstrate that local training is the primary contributor to energy consumption and carbon emissions, while communication overhead remains relatively minor. The experiments further show that optimizing aggregation and node selection strategies can effectively reduce the carbon footprint of DFL without significantly compromising model performance. Additionally, findings indicate that deploying models in regions with lower carbon intensity, leveraging early stopping mechanisms, and utilizing renewable energy sources can enhance the sustainability of DFL systems.

Despite these contributions, this study has certain limitations that provide avenues for future research. The proposed \textit{GreenDFL-SN} algorithm assumes that all nodes honestly report their energy consumption and carbon intensity, which may not always hold in real-world decentralized settings. Further research is needed to design incentive mechanisms or verification strategies to ensure reliable reporting. Additionally, this study estimates energy consumption based on hardware specifications and utilization metrics; incorporating real-time energy profiling can enhance accuracy. Future work can explore adaptive model selection strategies that dynamically adjust model complexity based on resource constraints and sustainability requirements. Moreover, integrating renewable energy-aware scheduling mechanisms and incentive models to encourage sustainable participation in DFL could further improve its environmental impact. Finally, extending the framework to account for dynamic carbon intensity profiles, such as time-of-day variations in renewable generation, represents an important avenue for future work. Larger-scale deployments across multiple countries could further enhance the analysis, although such settings face practical challenges related to data availability and deployment feasibility.

\section*{Declaration of Competing Interest}
The authors declare that they have no known competing financial interests or personal relationships that could have appeared to influence the work reported in this paper. 

\section*{CRediT authorship contribution statement}

\textbf{Chao Feng.} Methodology, Conceptualization, Writing - Review \& Editing.
\textbf{Alberto Huertas Celdrán.} Methodology, Writing - Review. 
\textbf{Xi Cheng.} Data curation, Analysis.
\textbf{Gérôme Bovet.} Project administration, Funding acquisition.
\textbf{Burkhard Stiller.} Supervision, Funding acquisition.

\balance
\section*{Acknowledgment}

This work has been partially supported by \textit{(a)} the Swiss Federal Office for Defense Procurement (armasuisse) with the CyberMind project (CYD-C-2020003) and \textit{(b)} the University of Zürich UZH.

\newpage

%
%

\bibliographystyle{cas-model2-names}
\bibliography{main}

\end{document}